RESEARCH PAPER

# High-time-resolution properties of 35 fast radio bursts detected by the Commensal Real-time ASKAP Fast Transients survey

D. R. Scott,[1] T. Dial,[2] A. Bera,[1] A. T. Deller,[2] M. Glowacki,[3,1,4] K. Gourdji,[5] C. W. James,[1] R. M. Shannon,[2] K. W. Bannister,[5] R. D. Ekers,[5,1] J. Paterson,[1] M. Sammons,[6,7] A. T. Sutinjo,[1] and P. A. Uttarkar[8]

[1]International Centre for Radio Astronomy Research, Curtin University, Bentley, 6102, WA, Australia
[2]Centre for Astrophysics and Supercomputing, Swinburne University of Technology, Hawthorn, 3122, VIC, Australia
[3]Institute for Astronomy, University of Edinburgh, Royal Observatory, Edinburgh, EH9 3HJ, United Kingdom
[4]Inter-University Institute for Data Intensive Astronomy, Department of Astronomy, University of Cape Town, Cape Town, South Africa
[5]CSIRO Space and Astronomy, Australia Telescope National Facility, PO Box 76, Epping, 1710, NSW, Australia
[6]Department of Physics, McGill University, 3600 rue University, Montréal, QC H3A 2T8, Canada
[7]Trottier Space Institute, McGill University, 3550 rue University, Montréal, QC H3A 2A7, Canada
[8]Centre for Astrophysics and Supercomputing, Swinburne University of Technology, John St., Hawthorn, VIC 3122, Australia
**Author for correspondence:** Clancy W. James, Email: clancy.james@curtin.edu.au.

## Abstract

We present microsecond-resolution, coherently-dedispersed, polarimetric measurements of 35 fast radio bursts (FRBs) detected during the Commensal Real-time ASKAP Fast Transients (CRAFT) incoherent sum (ICS) survey with the Australian Square Kilometre Array Pathfinder (ASKAP). We find a wide diversity of time–frequency morphology and polarisation properties broadly consistent with those of currently known non-repeating FRBs. The high S/N and fine time-resolution of our data however reveals a wealth of new information. Key results include (i) the distribution of scattering timescales, $\tau_{obs}$, is limited purely by instrumental effects, with no downturn at high $\tau_{obs}$ as expected from a log-normal distribution; (ii) for the 29 FRBs with known redshift, there is no detectable correlation between $\tau_{obs}$ and dispersion measure (DM) fluctuations about the Macquart relation, in contrast to expectations from pulsar scattering-DM relations; (iii) all FRBs probably have multiple components, and at least a large fraction have variable PA, the identification of which is limited by scattering; (iv) at least half of all FRBs exhibit PA microstructure at 200 μs–200 ns timescales, with behaviour most closely resembling a sub-category of Crab main pulses; (v) that there is a break in the FRB circular polarisation distribution at Stokes $V \gtrsim 20\%$, which is suggestive of a discrete sub-population.

## 1. Introduction

The mechanisms by which fast radio bursts (FRBs) — bright and unpredictable extragalactic bursts of radiation with durations typically of order milliseconds (Lorimer et al., 2007; Thornton et al., 2013) — are produced are not known. Characterising the FRB emission mechanism is made more complex by the great diversity in burst properties of the FRB population, including duration, spectral extent, time–frequency morphology, repetition, and polarisation (Pleunis et al., 2021), while propagation effects such as scattering can obscure the underlying FRB properties. Models attempting to explain FRB emission abound, and it is possible that there are multiple mechanisms that can produce FRBs (e.g. Platts et al., 2019). A general constraint is that the emission process must be a coherent one, deduced from the extreme brightness temperatures ($\gtrsim 10^{32}$ K, Lu & Kumar, 2019), far in excess of any known possible thermal process, and above what is allowed for synchrotron radiation (Kellermann & Pauliny-Toth, 1969).

The time–frequency morphologies of FRBs have been seen to be extremely varied, with some bursts containing one or more components of similar or varying frequency structure (e.g. Hessels et al., 2019; Day et al., 2020). Trends in morphology in bursts detected by the Canadian Hydrogen Intensity Mapping Experiment (CHIME) indicate a difference between repeating and non-repeating FRBs, with repeating bursts tending to be narrower in bandwidth and broader in duration (Pleunis et al., 2021).

Due to high data rates, most FRB searches are performed at millisecond time-resolution, which can integrate over detailed FRB structures. Studies at significantly higher time-resolution therefore offer great potential for discerning the nature of FRB emission as well as their environment (Farah et al., 2018). For example, Nimmo et al. (2022) observed in FRB 20200120E structure on timescales between 60 ns and 5 μs, which implies an upper limit on the emission region of 20–1500 m, ignoring relativistic beaming effects. This is inconsistent with emission from a synchrotron maser in a relativistic shock, instead favouring a magnetospheric origin. Hewitt et al. (2023) observe two distinct types of structure in FRB 20220912A: broad frequency-drifting components and extremely narrow "microshots", and propose that each arises from different emission processes.

Measured FRB polarisation properties also display diverse features and behaviours. These include high polarisation fractions and extremely large and variable rotation measures (RMs; Michilli et al., 2018); changes in the degree and basis of polarisation over the durations of bursts (e.g. Cho et al., 2020; Bera et al., 2025); depolarization at low frequencies (Feng et al., 2022); and swings in the polarisation position angle (PA; Luo et al., 2020; Mckinven et al., 2025).

While the majority of FRBs detected to date originate from a small number of powerful repeaters, it is important to determine the polarisation properties of the FRB population as a whole, especially since strongly repeating FRBs are known to have atypical spectro-temporal properties (e.g. Hessels et al., 2019). Sherman et al. (2024) recently presented high-time resolution polarimetric measurements of 25 apparently one-off FRBs detected by the Deep Synoptic Array (DSA). They found their sample to be consistent with all FRBs being intrinsically highly linearly polarised, with variations in the measured properties being due to effects of propagation through the media around the progenitors. Sand et al. (2025) analyse the high-time-resolution structure of 137 FRBs detected by the Canadian Hydrogen Intensity Mapping Experiment (CHIME) for which baseband data is available, including 12 repeat bursts, while Pandhi et al. (2024) analysed the polarisation properties of 128 non-repeaters. They find a consistent distribution of properties with those of DSA, identify sub-burst components as narrow as 23μs, and search for — but do not find — correlations between scattering timescale and other FRB properties. FRBs were classified according to the variability of their PA, and the apparent number of sub-components. Both analyses however identified signal-to-noise (S/N) and time resolution as limiting their ability to resolve FRB features. Given this, ongoing efforts to expand the sample of FRBs studied in full polarisation at high S/N and time resolution are strongly motivated.

The Australian Square Kilometre Array Pathfinder (ASKAP, Hotan et al., 2021) is a radio interferometer at Inyarrimanha Ilgari Bundara, the Murchison Radio-astronomy Observatory, that is applied to FRB searches by the Commensal Real-time ASKAP Fast Transients (CRAFT) collaboration. Through real-time detection & voltage capture (Bannister et al., 2019), polyphase filterbank (PFB) inversion (Morrison et al., 2020), and coherent beamforming, CRAFT is able to obtain coherently-summed measurements of the complex electric field of FRB signals in two orthogonal linear polarisations at the bandwidth-limited time resolution of $(336\,\mathrm{MHz})^{-1} \approx 3\,\mathrm{ns}$. This permits full polarimetric analysis at high-time resolution of FRBs detected with ASKAP, including one-off bursts for which the arrival times and positions are not known a priori. This was originally performed for FRB 20181112A by Cho et al. (2020), and has been applied to several particularly interesting ASKAP FRBs since.

A sample of 43 FRBs detected by ASKAP in incoherent sum mode (ICS; Bannister et al., 2019) has recently been presented by Shannon et al. (2025), 30 of which have high probability host galaxy associations and, hence, redshifts. Here, we present the sub-sample of 35 FRBs for which high-time-resolution data is available, of which 34 are consistently processed through the CRAFT effortless localisation and enhanced burst inspection (CELEBI) pipeline (Scott et al., 2023) to obtain extremely high-time resolution polarimetric data. §2 describes the methods of detection, data processing, and analysis, while §3 displays the bursts themselves and summarises their properties. In §4, we discuss the implications of this data, which we broadly divide into FRB classification (§4.1), scattering analysis (§4.2), polarisation analysis (§4.3), and other properties (§4.4). We summarise key results in §5.

We also note that the high-time resolution properties of a number of FRBs included in this sample have previously been studied. Day et al. (2020) presented FRBs 20180924B, 20190102C, 20190608B, 20190611B, and 20190711A, although these high time resolution data were obtained via an imaging method (with a maximum time resolution of 0.05 – 0.2 ms), rather than by PFB inversion and coherent beamforming. FRBs studied at high-time resolution using the CELEBI pipeline include FRB 20210912A (Bera et al., 2024), FRB 20210117A (Bhandari et al., 2023), and FRB 20230708A (Dial et al., 2025). The scintillation and scattering properties of nine of the bursts in our sample have also been previously studied (Sammons et al., 2023), while twelve have been analysed for depolarisation (Uttarkar et al., 2025). Unless otherwise noted, values have been re-derived to ensure a consistent — though not necessarily more accurate — sample, and so numbers reported may be discrepant between analyses. The primary exception to this is FRB 20181112A, where we take results from Cho et al. (2020).

## 2. Methods

### 2.1 Data acquisition

All FRBs presented here were detected using ASKAP's incoherent-sum detection mode (Bannister et al., 2019) to perform untargeted searches for bursts over a 336 MHz bandwidth, with a central frequency $\overline{\nu}$ varying between 832.5 MHz and 1631.5 MHz. This search is primarily performed commensally with other observations. Each of ASKAP's 12 m diameter antennas is fitted with a phased-array feed which forms 36 dual-polarisation beams over a total field of view of $\sim 30\,\mathrm{deg}^2$. The signal in each of these beams is integrated to a time-resolution of between 0.864–1.728 ms, incoherently summed across antennas, and searched for dispersed pulses in real time. Complex samples of the electric field in each beam of each antenna are stored in a ring buffer able to hold 3.1 s of 4+4 bit complex data in 336 1-MHz channels (since mid 2024, the bit resolution of the buffer has been changed to 1+1 bit in order to provide longer latency at the cost of increased quantisation noise). This is continuously written to until an FRB is detected, at which point these voltage buffers are frozen and downloaded from all available antennas for further processing. See Shannon et al. (2025) for a fuller description of the history of this observation mode.

### 2.2 Processing

We process the raw voltages from each FRB to obtain high-time resolution polarimetric data using CELEBI (Scott et al., 2023), updates to which will be described in a future work (Glowacki et al., in prep.). In short, after interferometric burst localisation, CELEBI performs polyphase filterbank inversion and beamforming to obtain the complex electric fields in two orthogonal linear polarisations (which, following the terminology used in Hotan et al. (2021), we refer to as X and Y) at the bandwidth-limited time resolution of $(336\,\mathrm{MHz})^{-1} \approx 3\,\mathrm{ns}$, coherently summed across all antennas for which data is available. These data are initially coherently dedispersed to the DM measured at detection.

Corrections for polarisation leakage — calculated from observations of a polarisation calibrator accounting for the residual instrumental X–Y phase lag and ellipticity — are applied to the X and Y voltage data. Complex voltage dynamic spectra in X and Y are then constructed with 336 channels of width 1 MHz and time resolution 1 μs via a 336-point complex-to-complex fast Fourier transform. Dynamic spectra in each of the Stokes parameters I, Q, U, and V are derived directly from the X and Y data.

We note that five of the FRBs presented in this paper — FRBs 20210320C, 20211212A, 20220725A, 20240201A, 20240210A — were detected significantly off-axis in the edge/corner beams of the ASKAP phased array feed (see Hotan et al., 2021, for details), i.e. outside the half power point of the respective beam across the observing frequency band. The measured polarisation properties of these FRBs may be affected by complex and uncorrected instrumental response even if polarisation calibration has been applied, as the calibration solutions — calculated with the calibrator at the centre of the respective beam — may not be adequate to correct for the instrument-response at the position of the FRB. A preliminary investigation suggests that, in the worst-case scenario where the FRB location coincides with a null in the (frequency-dependent) beam pattern, relative systematic errors $\Delta Q/Q$, $\Delta U/U$, and $\Delta V/V$ could be as large as 30%. For the rest of our sample, in the worst-case scenario of an FRB detected at the half-power point on the outer edge of a corner beam, the error is expected to be at most 10%. We intend to re-examine these FRBs in a future work, and for now, we mark these five off-axis FRBs in our data tables, and suggest caution when using their polarisation properties.

#### 2.2.1 Bad channel removal

Parts of the observing band of some FRBs are affected by RFI, which are identified using the off-pulse noise statistics and masked. The remaining channels of the dynamic spectra are then normalised such that the off-pulse noise has a mean of zero and standard deviation of unity (assuming Gaussian noise).

The 3.1 s duration of the voltage buffers and the latency of the real-time detection sometimes ($\mathcal{O} \sim 2\,\mathrm{s}$) results in some of the higher-frequency components of the burst falling out of the ring buffers before they can be frozen and the data downloaded due to the dispersive time delay, which can be $\sim 4\,\mathrm{s}$ for a $1{,}000\,\mathrm{pc\,cm^{-3}}$ FRB observed at the lower end of ASKAP's frequency range. We therefore remove these channels in order to avoid adding unnecessary noise to the time profiles, and to prevent the false impression of a sharp spectral cutoff in the dynamic spectra.

Once channels have been removed, we integrate the dynamic spectra in all four Stokes parameters over frequency to obtain the respective time series profiles.

#### 2.2.2 Dedispersion

We perform a structure-optimising dedispersion using the method of Sutinjo et al. (2023), which maximises the sum of $\left(d\tilde{I}/dt\right)^2$. $\tilde{I}$ is the FRB's Stokes-I profile (constructed using the above procedure) after being averaged in time ("t-scrunched") and passed through a low-pass filter to remove noise. The filter frequency is chosen to obtain the lowest uncertainty on the structure-maximising DM, $\mathrm{DM_{struct}}$. All results shown in this work have been coherently dedispersed to $\mathrm{DM_{struct}}$.

### 2.3 Analysis

The following analyses are performed on high-time-resolution, dedispersed dynamic spectra. In what follows, we calculate errors on quantities by measuring the off-pulse standard deviation in the Stokes parameters, $\sigma_{I,Q,U,V}$, and applying standard error propagation techniques.

#### 2.3.1 Signal-to-noise

Relative to our real-time ICS detection system, our offline S/N is boosted due to coherent beamforming, coherent dedispersion, better RFI flagging, and higher time resolution. However, for high-DM FRBs, our real-time system is sometimes unable to generate a trigger in time to capture the entire FRB signal in our buffer, resulting in the loss of high-frequency components, and degrading the S/N.

We calculate our offline S/N values, $\mathrm{S/N_{off}}$, using a moving boxcar of variable width, and report the maximum value over all possible boxcar windows. For the sample presented here, these effects result in S/N of between 21 and 479, representing no gain, to a factor of 15 gain, above the real-time detection S/N.

#### 2.3.2 Width

The apparent width of an FRB is an ill-defined quantity, with many FRBs exhibiting broad structures that only appear at low S/N. Hence, any measure of FRB width will be dependent on the S/N at which that FRB is detected. The S/N-maximising width calculated above typically reflects only the strongest, narrow peak of an FRB, and ignores broader underlying structures.

Instead, we choose to define a box-car width $w_{95}$ as the smallest width that encloses 95% of the total burst fluence. We also defined the zero point reference in time to bisect the fluence distribution.

#### 2.3.3 Rotation measure

To fit for rotation measure (RM), we first integrated the Stokes $I$, $Q$ and $U$ dynamic spectrum of each FRB over the burst window to obtain 1 MHz Stokes spectra. We employed two different techniques. QUfitting was used as the primary method for its robustness against band-limited FRBs such as FRB 20211212A and FRB 20220610A. Our implementation of

QUfitting is a Bayesian method that maximises the likelihood (Bannister et al., 2019, Supplementary material)

$$\mathcal{L} = \prod_{i=1}^{N} \frac{1}{\sqrt{4\pi^2\sigma_Q^2\sigma_U^2}} \exp\left(\frac{-\left[Q_i - L_i'\cos(2\psi(\nu_i))\right]^2}{2\sigma_Q^2}\right) \times \exp\left(\frac{-\left[U_i - L_i'\sin(2\psi(\nu_i))\right]^2}{2\sigma_U^2}\right), \tag{1}$$

where $L'$ is the de-biased total linear polarisation (Everett & Weisberg, 2001),

$$L' = \begin{cases} \sigma_I\sqrt{\left(\frac{L}{\sigma_I}\right)^2 - 1} & \frac{L}{\sigma_I} > 1.57 \\ 0 & \text{otherwise,} \end{cases} \tag{2}$$

$$L = \sqrt{Q^2 + U^2}, \tag{3}$$

and $\psi$ is the polarisation position angle (PA),

$$\psi(\nu_i) = \mathrm{RM}c^2\nu_i^{-2} + \psi_0. \tag{4}$$

Here, $\sigma_{I_i}$, $\sigma_{Q_i}$ and $\sigma_{U_i}$ are measured as the off-pulse noise in the Stokes *I, Q* and *U* dynamic spectra in channel *i* with frequency $\nu_i$. RM and $\psi_0$ are sampled using Dynesty (dynamic nested sampling; Speagle, 2020) within Bilby (Ashton et al., 2019).

Measurements of RM require properly calibrated polarisation, which is achieved for a majority of FRBs in our sample. However, there are two situations in which we are unable to derive adequate solutions for polarisation calibration. (1) J1644-4559 is used as a calibrator at the ASKAP low-band (central frequency $\leq$ 919.5 MHz). Its period of 0.455 s is not much shorter than the 3.1 s voltage download, meaning that only a small number of pulses can be captured. With the addition of high scattering at low frequency, there is not enough S/N to properly model and remove any polarisation leakage. (2) The FRB was detected on an edge/corner beam where there is potentially high polarisation leakage that we can not model appropriately with our current polarisation calibration. For the sub-sample of FRBs that do not meet the conditions for polarisation calibration, we can still estimate the RM using Faraday rotation measure synthesis (Brentjens & de Bruyn, 2005) as long as the true integrated RM is not close to zero. This is possible since we expect any polarisation leakage to manifest as a peak centered at zero in the Faraday depth function (FDF) (Ng et al., 2020).

To estimate the RM using the FDF, we first mask any channels where the Stokes power $I(f) < 3.0\sigma_I$, which is significant in the case of band-limited and/or highly scintillating FRBs. The masked Stokes spectra and off-pulse channel variance were then passed to RM-Tools (Purcell et al., 2020).

Of our two methods, FDF is the most robust, and produces results for all FRBs analysed. For FRBs where both QUfitting and FDF results are available, the estimated RM agrees within the errors for all except one FRB (20211212A), which is sufficiently band-limited that we do not consider either estimate reliable.

Using our best-fit RM, we correct $Q$ and $U$ according to $\psi(f)$ from the best-fit RM,

$$\begin{aligned} Q'(t,f) &= Q(t,f)\cos(2\psi(f)) + U(t,f)\sin(2\psi(f)), \\ U'(t,f) &= -Q(t,f)\sin(2\psi(f)) + U(t,f)\cos(2\psi(f)), \end{aligned} \tag{5}$$

and integrate over frequency to obtain time-profiles.

We also estimate the Galactic RM contribution $\mathrm{RM_{MW}}$ along the line-of-sight to the FRB source as given by Hutschenreuter et al. (2022) using the `FRB` software package (Prochaska et al., 2023), and subtract this from the total RM to determine the extragalactic RM contribution, $\mathrm{RM_{EG}}$.

In this work, we do not investigate any change in RM over time within a burst.

#### 2.3.4 Position angle

To calculate burst PA profiles, we follow the method outlined in Day et al. (2020). To briefly summarise, we calculate the PA $\psi(t)$ as

$$\psi(t) = \frac{1}{2}\arctan\left(\frac{U'(t)}{Q'(t)}\right), \tag{6}$$

where $Q'(t)$ and $U'(t)$ are the RM-corrected Stokes Q and U profiles. We then plot only time steps where the total linear polarisation fraction is significant, i.e. $L' > 3\sigma_I$.

#### 2.3.5 Polarisation fractions

Average polarisation properties are calculated by integrating $I(t)$, $L'(t)$, and $|V(t)|'$ profiles over the burst window $w_{95}$ to obtain $\bar{I}$, $\bar{L}$, and $|\bar{V}|$, and corresponding polarisation fractions, $\bar{L}/\bar{I}$ and $\bar{V}/\bar{I}$. $|V(t)|$ is de-biased to $|V(t)|'$ to account for the modulus (Karastergiou et al., 2003; Oswald et al., 2023),

$$|V|' = \begin{cases} |V| - \sigma_I\sqrt{\frac{2}{\pi}}, & |V| > \sigma_I\sqrt{\frac{2}{\pi}} \\ 0 & \text{otherwise.} \end{cases} \tag{7}$$

We also calculate the total polarisation fraction

$$\frac{P}{I} = \frac{\sqrt{\bar{L}^2 + |\bar{V}|^2}}{\bar{I}}. \tag{8}$$

#### 2.3.6 PA swings

We characterise the position angle (PA) of our FRBs into three classes: those exhibiting constant PA, those showing a linear trend, and those with a variable PA. To do this, we use an F-test to determine if the reduction in $\chi^2$ from fitting polynomials of higher order provides a significantly better fit than those of lower order. While quadratic and cubic polynomials generally do not reflect the actual behaviour of FRBs with non-linear PA swings, these fits are sufficient to identify the overall trend. We use a p-value threshold of 0.01 to reject a lower-order polynomial in favour of a higher-order polynomial (linear vs. constant; quadratic or cubic vs. linear or constant). We also calculate the $\chi^2$ for each fit to determine if the fits are good.

A fuller analysis of FRB polarisation state variation is best conducted using the Poincare sphere (e.g. Cho et al., 2020;

Bera et al., 2025). Here, we do not perform such a detailed analysis for each burst, and instead, attempt to characterise each FRB as being consistent with constant, linear, or variable (non-linear) PA variation.

#### 2.3.7 Scattering

Passage through turbulent media in the Milky Way, intergalactic medium, FRB host galaxy, and/or in the vicinity of the FRB progenitor, produces multi-path scattering in FRBs. In the case of a single effective scattering screen, this manifests as both an exponential smearing of the time-profile with timescale $\tau(\nu) \propto \nu^{\alpha}$, and a scintillated frequency spectrum (see §2.3.8). Due to the complicated sub-pulse structure of FRBs however, it can be difficult to distinguish scattering from the intrinsic pulse shape. Furthermore, we wish to avoid the assumption of $\alpha = -4$ or $\alpha = -4.4$ which is expected from Gaussian or Kolmogorov turbulence: many pulsars exhibit smaller values of $\alpha$, as high as $-1.5$ (Bhat et al., 2004; Geyer et al., 2017; Krishnakumar et al., 2017), and there is no guarantee that FRBs will behave any differently.

To determine the scattering time and test the robustness of our fitting results, we divide the bandwidth in which each FRB has significant power into four sub-bands. We first fit these independently with a sequence of N Gaussian burst profiles (each defined by a width, central time, and amplitude), and an exponential scattering term $\tau$. We consider all values of $N$ where an F-test returns the chance probability of a significant decrease in $\chi^2$ by adding successive Gaussian components to be 0.01 or less. However, we also find that such an approach can lead to over-fitting, and very large values of $N$ that may not necessarily reflect the true structure.

We therefore also evaluate goodness-of-fit by testing if the values of $\tau$ from well-fitting models follow a power-law, $\tau(\nu) \propto \tau_{\rm obs}(\nu/\overline{\nu})^{\alpha}$, where $\overline{\nu}$ is the mean fitted frequency of the FRB. We check the robustness of this method by enforcing such a power-law dependence across the four frequency bands in the Gaussian fitting procedure, simultaneously fitting a single value of $\tau_{\rm obs}$ and $\alpha$ to all four bands. We then scale scattering results to a standard frequency of 1 GHz, $\tau_{1\,\rm GHz}$, using the power-law fits.

We have not, however, been able to construct a fully automated fitting method that returns reliable results with appropriate errors in all cases. We therefore evaluate the reasonableness of model fits by eye, and quote error ranges in $\tau_{\rm obs}$, $\tau_{1\,\rm GHz}$, and $\alpha$ to cover the range of reasonable models, as well as the statistical errors returned by the fitting procedure. This also means that the value of $\tau_{1\,\rm GHz}$ and its uncertainty will only be approximately consistent with $\overline{\nu}$, $\tau_{\rm obs}$, and $\alpha$.

We hope to improve our scattering fit procedure in the future.

#### 2.3.8 Scintillation

For each burst, we calculate the spectral modulation index ($m$) by applying a similar method to that outlined in Sammons et al. (2023), on 0.1 MHz resolution Stokes I spectra, constructed according to the above procedure, integrated over the duration of the burst. To focus on scintillation-scale fluctuations, we again divide the entire bandwidth into four sub-channels and calculate $m$ as the mean of the modulation indices in each sub-channel. In bursts with substantial small-scale modulation ($m \geq 0.4$), we fit for the decorrelation bandwidth $\nu_{\rm DC}$ in each sub-channel and the scintillation spectral index across all sub-channels following the methods of Sammons et al. (2023).

## 3. Results

We consider the sample of 43 ASKAP FRBs detected in incoherent sum (ICS) mode reported by Shannon et al. (2025), spanning 24 September 2018 to 18 March 2024. Of these, no suitable voltage data was captured for six of them, preventing offline analysis. In general, this was because either the voltage download was not triggered (bursts detected above a given width threshold were not triggered due to challenges with false positives due to RFI), or the voltage download did not complete correctly (on some occasions, only a single polarisation was downloaded, or only a small subset of the frequency channels and/or antennas completed the download). A special case is FRB 20190714A, where only one polarisation product was downloaded, preventing full polarisation products from being derived. Additionally, FRB 20181112A was excluded from reprocessing with CELEBI due to an incompatibility of the data format with the current version of the pipeline, but we quote properties measured from the dedicated high-time resolution analysis by Cho et al. (2020) where possible. This leaves us with 34 bursts in our fully-processed sample to date (five of which may have polarisation calibration errors), and one further FRB with partial results (FRB 20181112A).

Figure 1 is a gallery of these 35 FRBs. The time resolution for each FRB has been chosen in order to display the peak of each burst with S/N$\sim$ 20. Tables listing the observational parameters and measured burst properties of each FRB — include properties of FRB 20181112A from Cho et al. (2020) where applicable — are given in the Appendix.

## 4. Discussion

In the following discussion and analysis, we include data from all FRBs for which high-time-resolution data is available, i.e. the full sample of 36.

### 4.1 Classification

Previous classifications of FRBs have focused on both their time-frequency structure, with features such as number of components and spectral occupancy, and also their fractions of linear and circular polarisation (Pleunis et al., 2021; Sherman et al., 2024; Sand et al., 2025). The two most comparable FRB samples to that presented here — 25 FRBs detected by the Deep Synoptic Array (DSA; time resolution 32.768 μs, Sherman et al., 2024), and 128 non-repeating FRBs detected by the Canadian Hydrogen Intensity Mapping Experiment (CHIME/FRB; time resolution 2.56 μs, Pandhi et al., 2024) — are both derived through offline analysis of voltage data, similarly to our CELEBI pipeline. Key differences are that our gain in offline S/N is greater than that of both, due to our

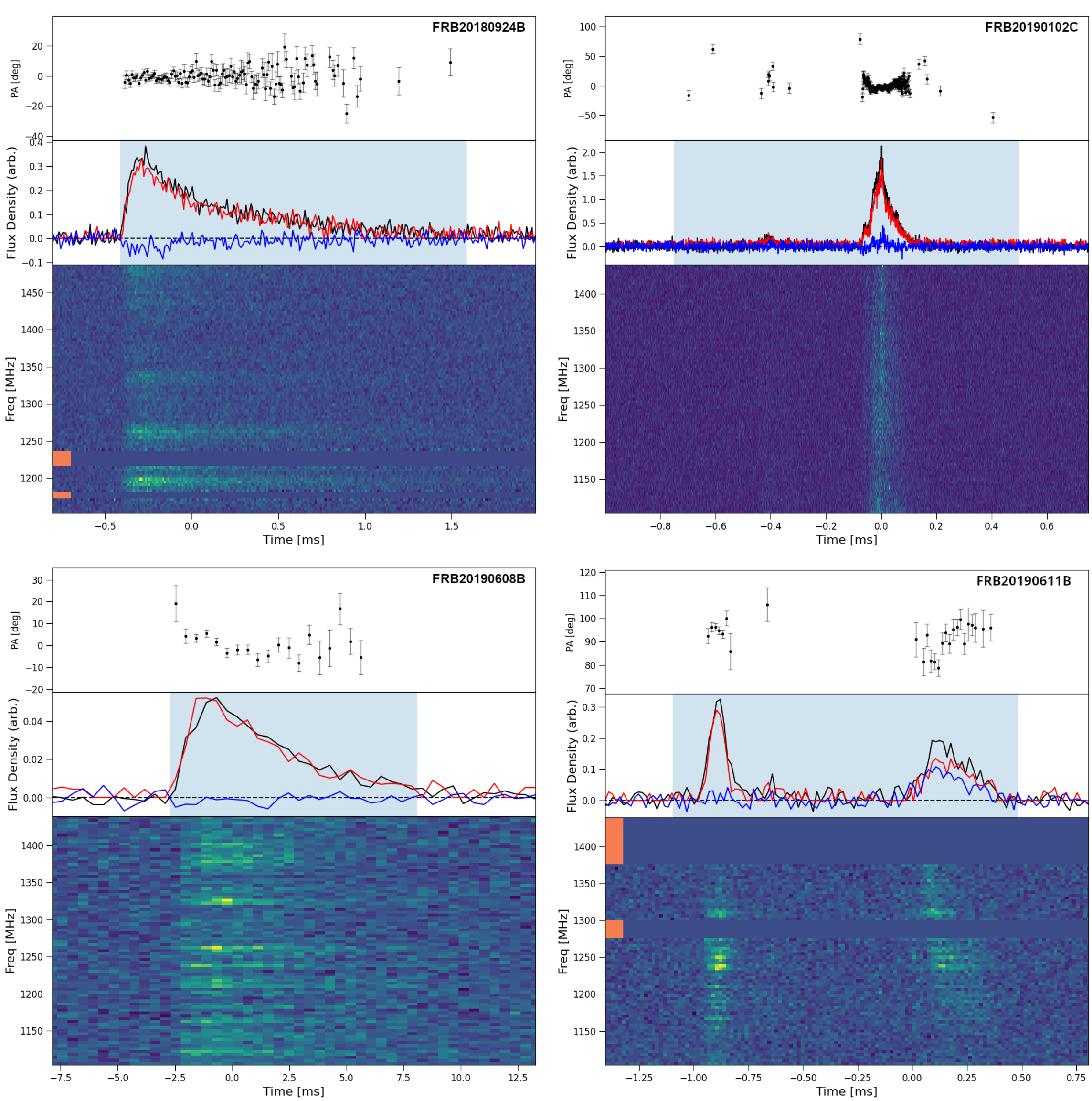

**Figure 1.** High-time resolution dynamic spectra and polarimetric profiles of the FRBs in our sample. The zero-point in time is set relative to the bin with the peak I profile. Top panels in each figure: PA relative to the value at the peak time index ($\Delta\psi = \mathrm{PA}(t) - \mathrm{PA}(0)$). Middle panels: profiles of total intensity $I$ (black), and bias-corrected linear polarisation $L'$ (red) and circular polarisation $|V|'$ (blue), with the region corresponding to the optimal fitted boxcar of width $w_{95}$ containing 95% of the fluence shaded in blue. Bottom panel: Stokes I dynamic spectrum with frequency resolution 4 MHz. Frequencies above the crossing frequency (where the FRB has fallen off the edge of the voltage buffer due to its dispersive sweep) are denoted in pink, while regions dominated by RFI are denoted in orange; both are set to zero. All bursts have been coherently dedispersed to the DM indicated in Table 1, and corrected for Faraday rotation by their respective RM.

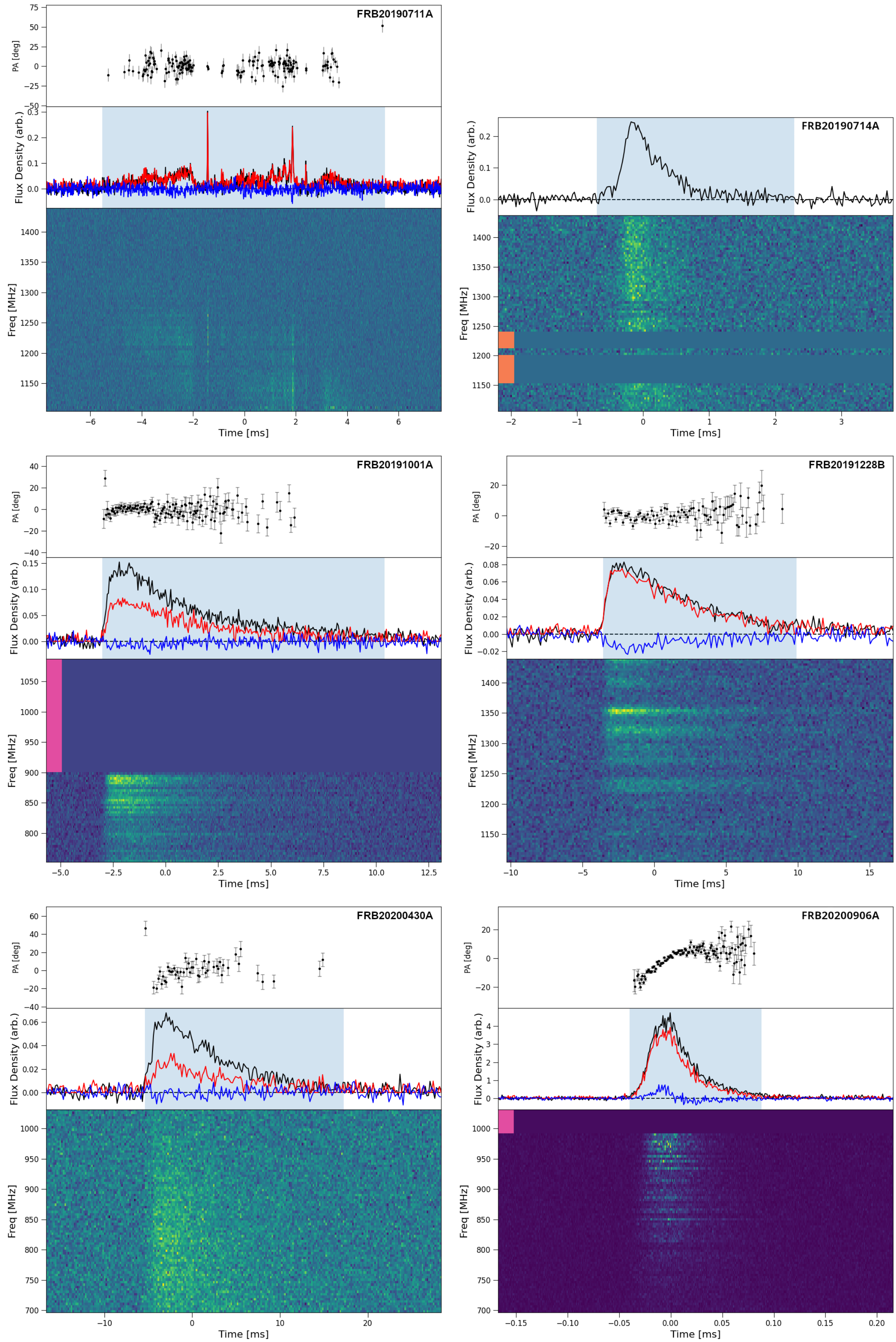

**Figure 1.** (cont.) FRB 20190714A is missing Y polarisation data, and hence, only its Stokes $I$ time profile using X polarisation can be calculated, and no PA data is available.

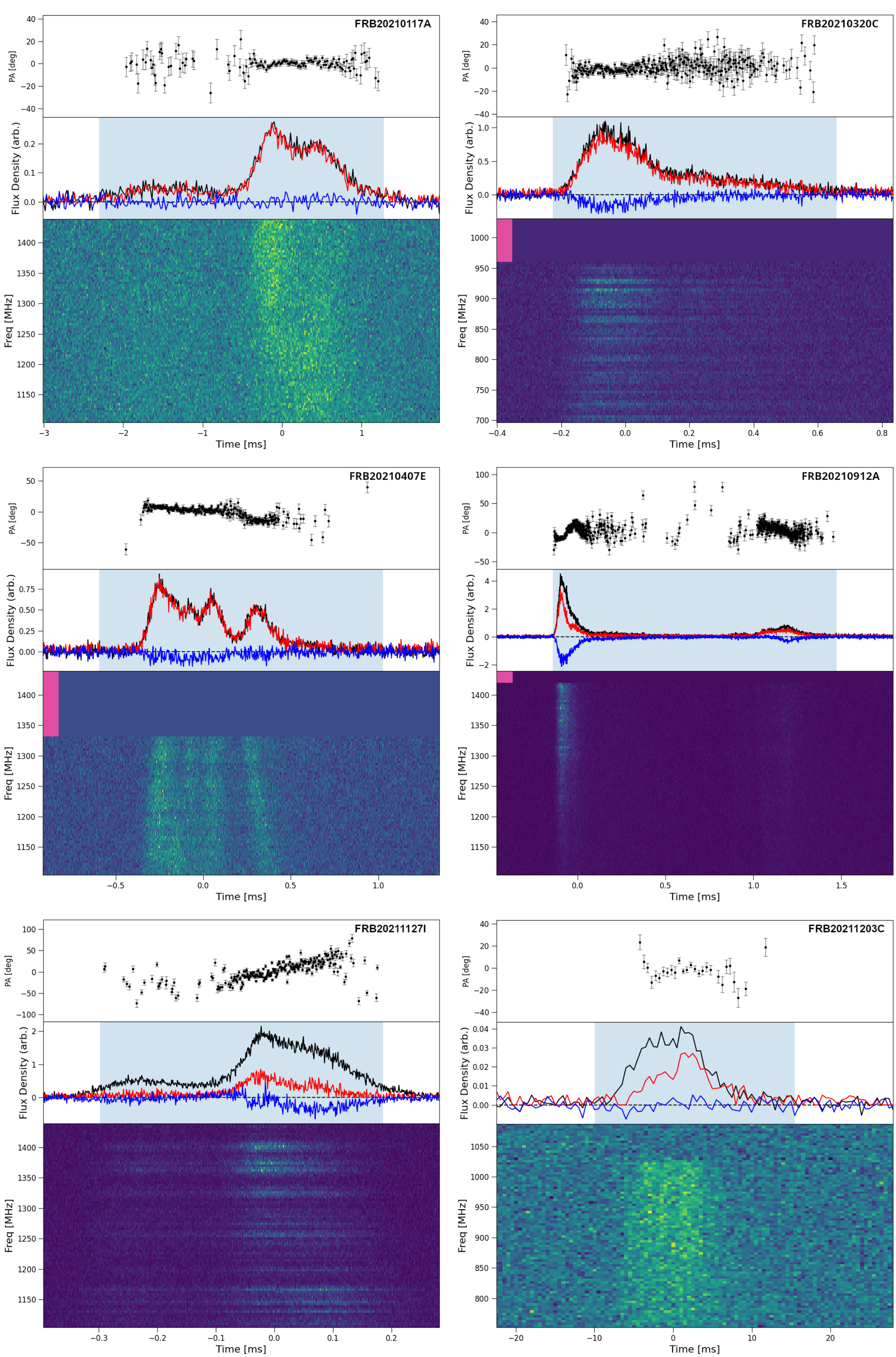

**Figure 1.** (cont.)

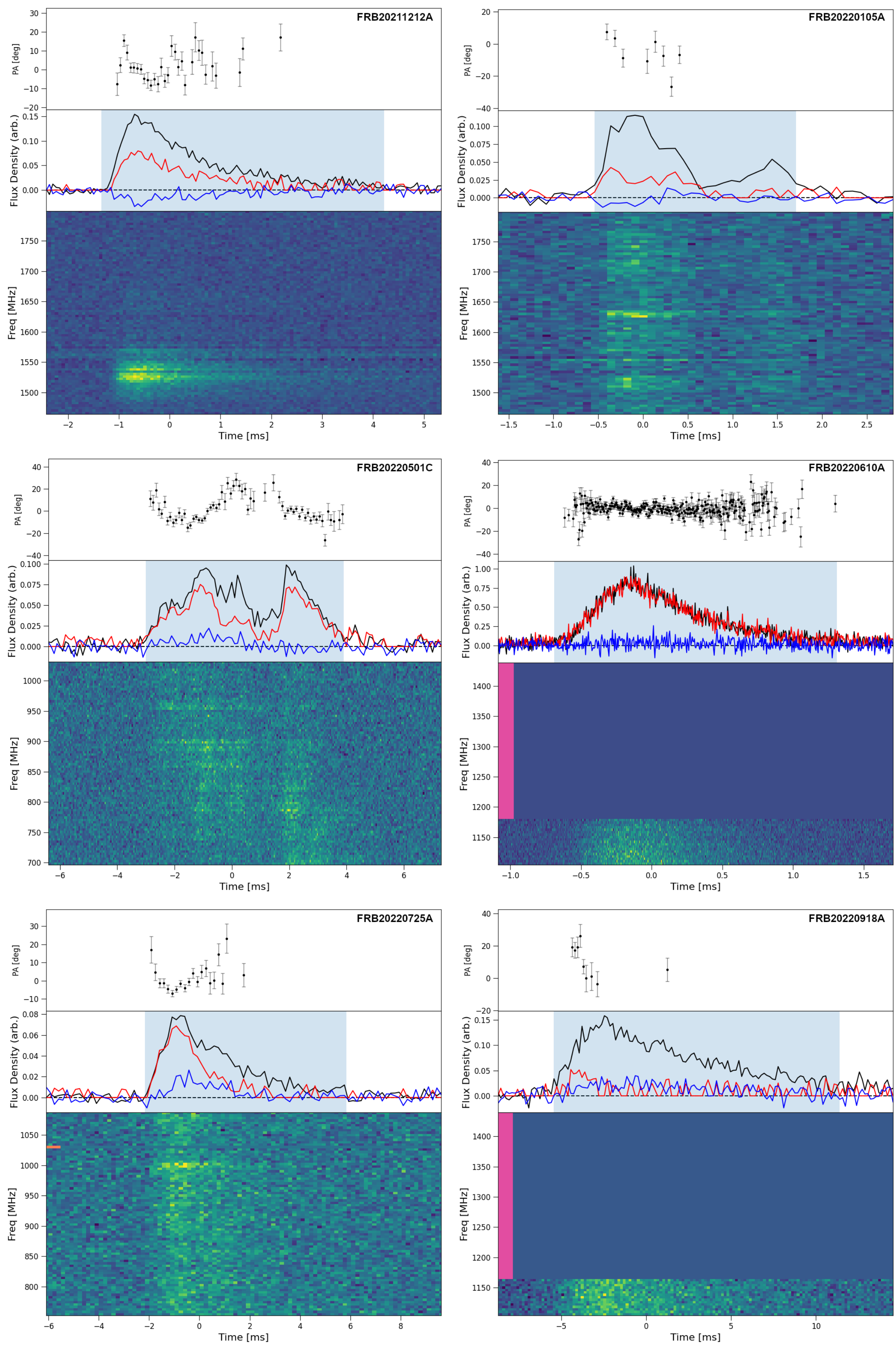

**Figure 1.** (cont.)

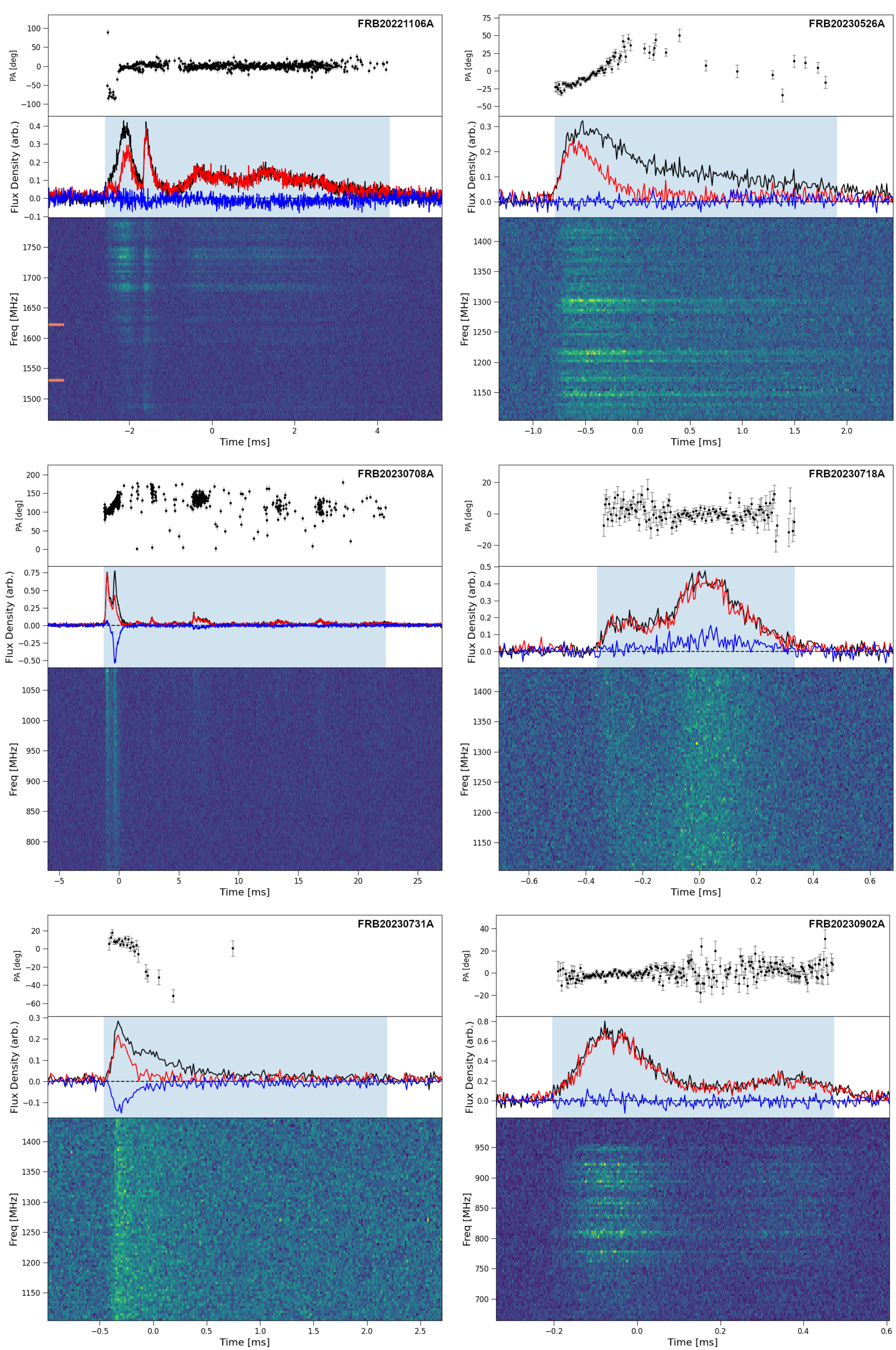

**Figure 1.** (cont.)

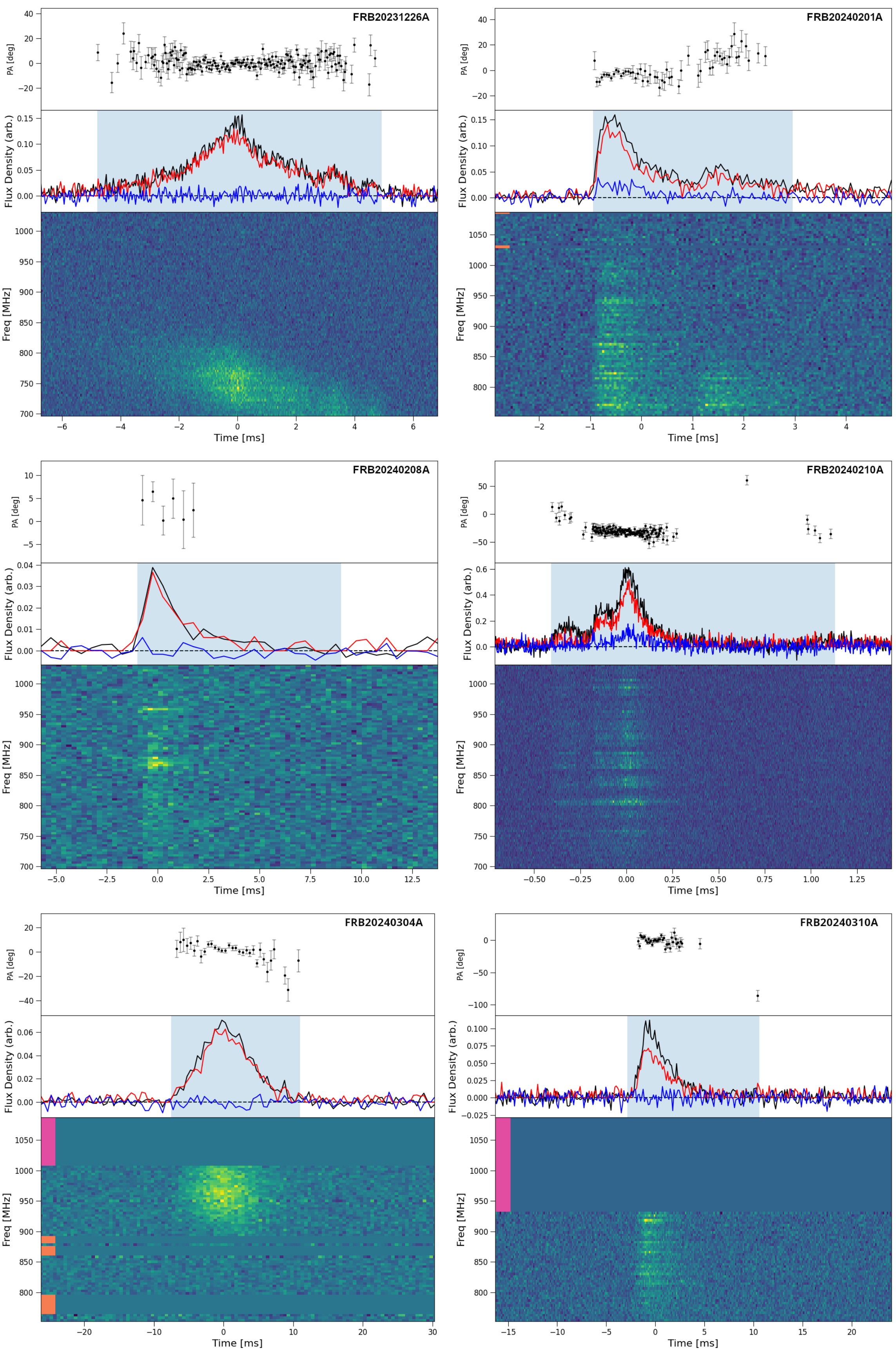

**Figure 1.** (cont.)

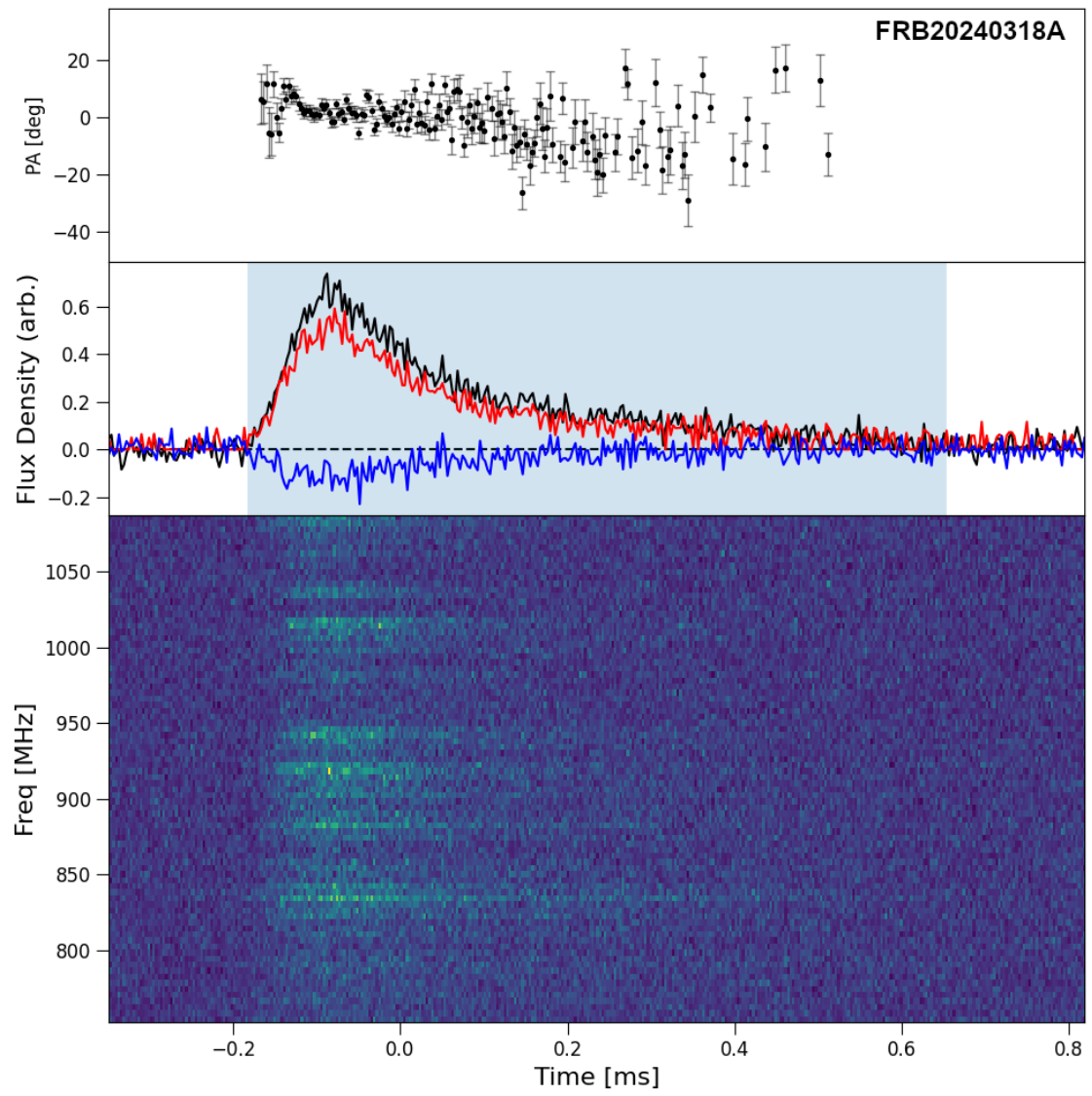

**Figure 1.** (cont.)

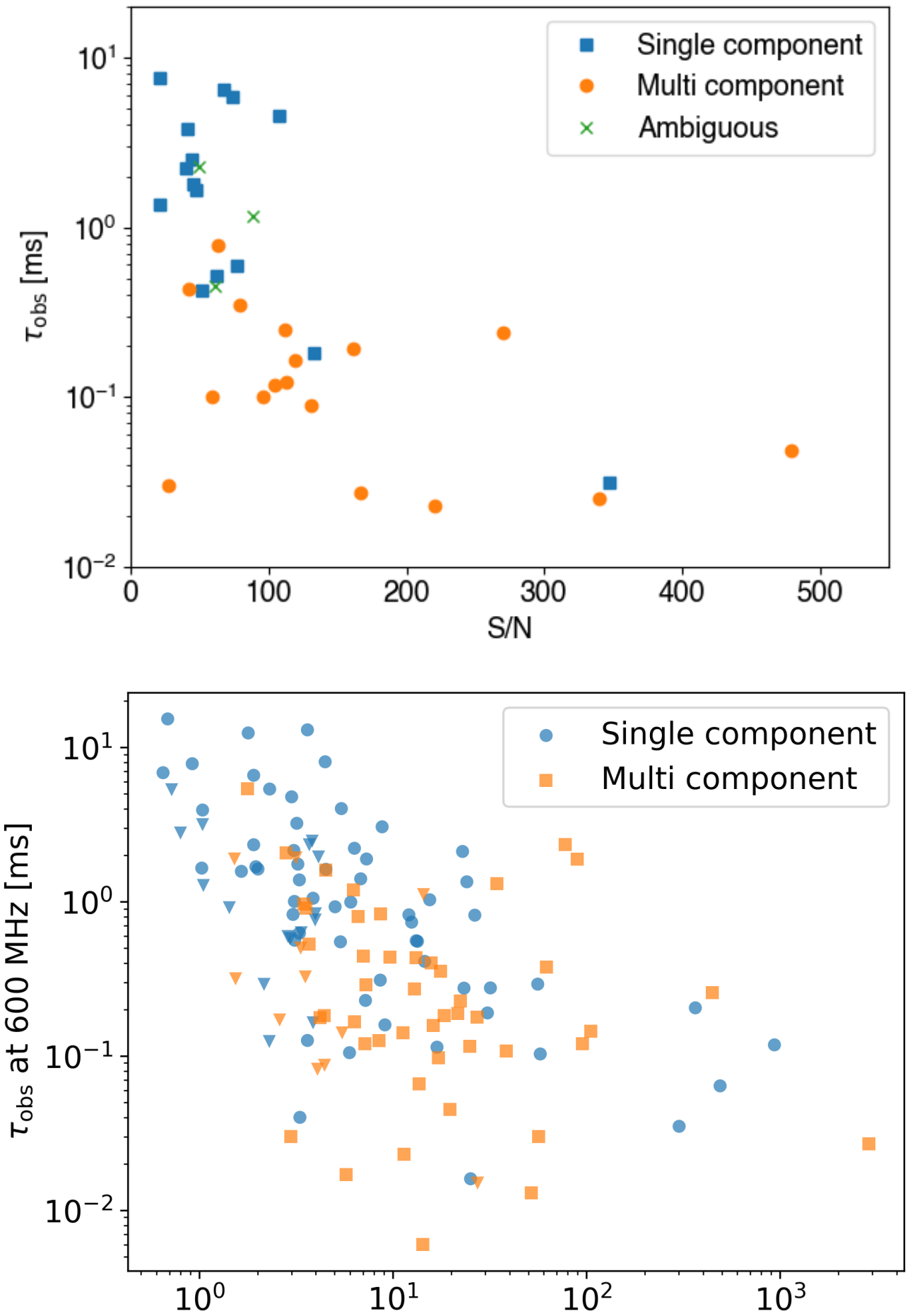

**Figure 2.** Scattering at central frequency, $\tau_{obs}$, as a function of (top): offline signal-to-noise ratio S/N for the ASKAP HTR sample presented here, and (bottom) the CHIME baseband catalog from CHIME/FRB Collaboration et al. (2024), for bursts with single and multiple identifiable components. Peak flux is used for the latter, as S/N information is not included in that catalog. Scattering upper limit measurements are denoted by triangular markers.

incoherent online detection system; and that CHIME do not consider their circular polarisation measurements reliable, and only publish values of $L/I$. These samples have been morphologically classified into those with single/multiple components, and those with constant and variable PA; and they are also divided according to unpolarised ($L/I < 0.35, |V|/I < 0.3$), partially polarised ($0.35 < L/I < 0.7, |V|/I < 0.3$), linearly polarised ($L/I > 0.7, |V|/I < 0.3$), and circularly polarised ($|V|/I > 0.3$) bursts.

As shown in Figure 2 however, the FRBs we identify as being single-component tend to have high scattering values, with only one (FRB 20200906A) having a scattering value below 0.1 ms. Since we expect scattering to originate in media far from the source, this cannot be an intrinsic effect, and suggests that all FRBs would be detected to have multiple components, which may not always be identified due to time resolution, scattering, and S/N constraints. We also show the same effect for CHIME FRBs with baseband data (CHIME/FRB Collaboration et al., 2024). Although the distinction in that sample is not so clear, there is an obvious trend showing that the fraction of FRBs with multiple components is strongly anti-correlated with scattering timescale.

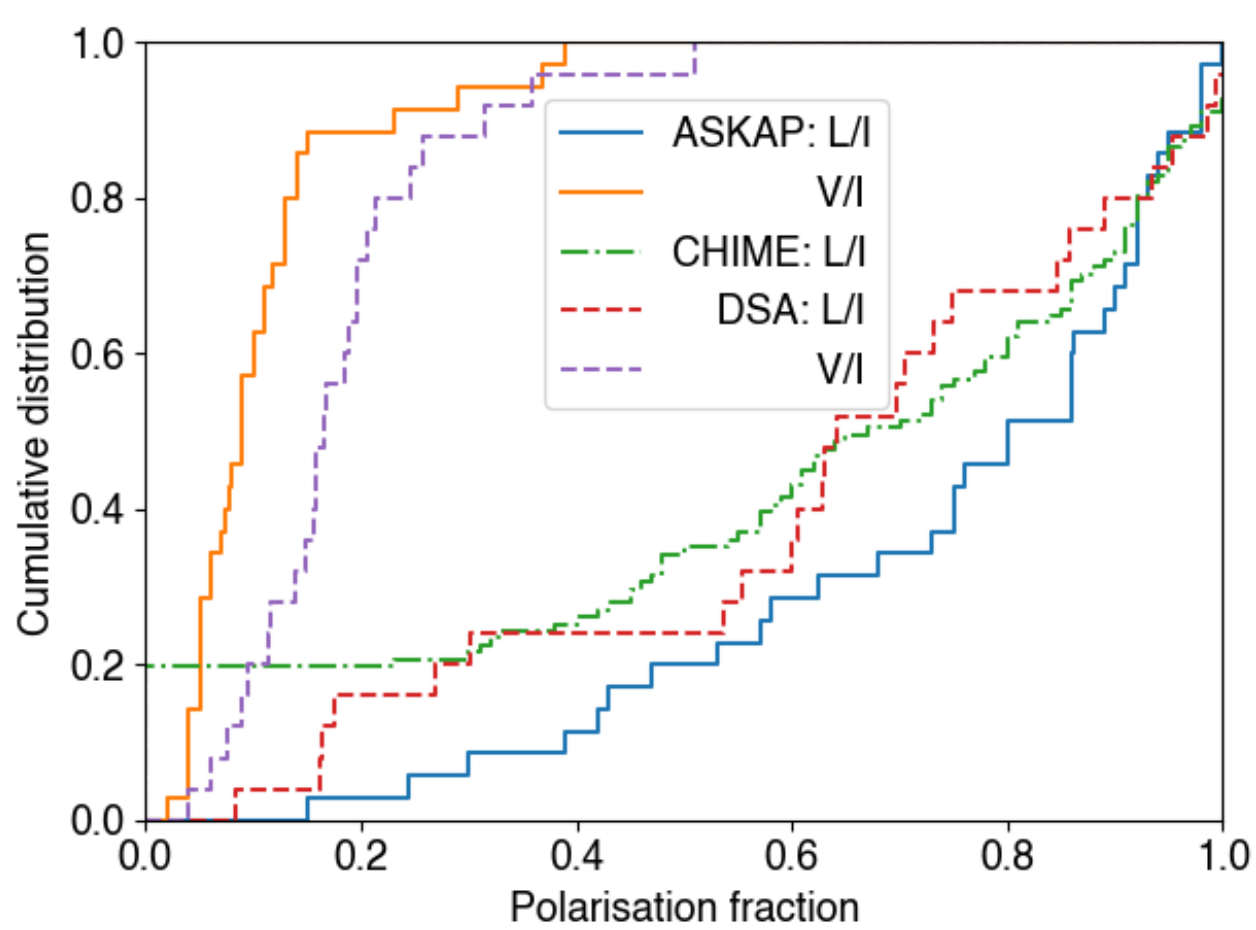

**Figure 3.** Cumulative distributions of total linear (L/I) and circular (V/I) polarisation fractions from ASKAP ICS observations (this work), CHIME baseband data (Pandhi et al., 2024) and DSA (Sherman et al., 2024). Note that V/I values are not available for CHIME.

Cumulative histograms of polarisation fractions from ASKAP, DSA, and CHIME FRBs are shown in Figure 3. We find some evidence for a higher linear polarisation fraction than both CHIME (p-value of 17% on a 2-sample Kolmogorov-Smirnov test; Kolmogorov, 1933; Smirnov, 1948) and DSA (7%) samples, and a significantly lower circular polarisation fraction than DSA (0.04%).

The absolute amount of circular polarisation we observe is lower than that found by DSA. The cause of this may be due to frequency dependence in the sign of $V$ which, when integrated over ASKAP's larger bandwidth, results in a reduced total $V$. We leave an investigation of this, and other frequency-dependent polarisation effects, to a future work.

We do, however, observe the same break in the cumulative distribution of $V/I$, which occurs at $V/I \sim 15\%$ for our data, and at $V/I \sim 25\%$ in DSA data. This suggests a distinct sub-class of highly circularly polarised FRBs, consisting of $\sim 10\%$ of the population with $V/I > 0.2$; in contrast, there is no clear distinction between strongly and weakly linearly polarised FRBs. This result is robust against polarisation calibration errors: we have checked that the distributions of $L/I$ and $V/I$ from FRBs calibrated with Vela are consistent with that from other calibrators, with two-sample KS-tests showing that differences between cumulative distributions of $L/I$ and $V/I$ at least as large as that observed would be expected 16% and 92% of the time under the null hypothesis that these calibrators perform equally well.

We therefore conclude that our highly circularly polarised FRBs may be an intrinsic sub-class of the FRB population, and we encourage investigations as to whether this reflects fundamentally different progenitors, emission mechanisms, or lines of sight through a nominal neutron star progenitor's magnetosphere. In particular, we note that relativistic magnetised plasmas can induce circular polarisation in FRBs (Kumar et al., 2023, e.g.) — if this is the cause, the question then becomes why some FRBs show evidence for such plasmas, and others do not. We leave such investigations to future works.

We also conclude that classifying FRBs according to single- and multi-component bursts, or the fraction of linear polarisation, is not a meaningful distinction.

### 4.2 Scattering

Our multi-component scattering fits produced mixed results. Of our 36 FRBs, 15 produced good scattering fits, which we define as errors of less than 10% on $\tau_{\rm obs}$, and an error of less than unity on $\alpha$. These are typically FRBs with large scattering times. Eight produced poor fits, which are readily identifiable as those FRBs where $\alpha$ is consistent with zero at the $1\sigma$ level. This is due to those FRBs having relatively low values of scattering, and profiles where scattering is difficult to distinguish. They do not necessarily have large errors on $\tau_{\rm obs}$ however, since these errors reflect the range of values given by plausible fits as described in §2.3.7 — and such fits may consistently converge to the same incorrect answer.

Our cumulative distribution of scattering values is plotted against that from CHIME/FRB Collaboration et al. (2024) in Figure 4. When we make no adjustment for CHIME having a lower frequency, the two distributions are very similar. Scaling CHIME's observed values of $\tau$ to 1 GHz by the expected factor of $(600\ \text{MHz}\ /1\ \text{GHz})^4 \sim 0.13$ however would make CHIME's and ASKAP's observations highly discrepant. We explain this as both distributions being dominated by selection effects, rather than the intrinsic distribution: for $\tau > 2$ ms, scattering begins to dominate the observed FRB duration in each experiment, reducing the number of highly scattering FRBs detected.

CHIME/FRB Collaboration et al. (2021) have used pulse injections to correct for selection effects in their data, and modelled the intrinsic scattering timescale at 600 MHz using a log-normal distribution, $\log(\tau/1\ \text{ms}) \sim N(\mu_\tau, \sigma_\tau)$, finding $\mu_\tau = 0.7$, and $\sigma_\tau = 1.72$ (CHIME/FRB Collaboration et al., 2021). However, their selection-effect-corrected histogram of scattering timescales (the right-hand panel of Figure 17 of their paper) is at least equally consistent with a constant distribution of scattering timescale per log bin above 1 ms, and our measurements.

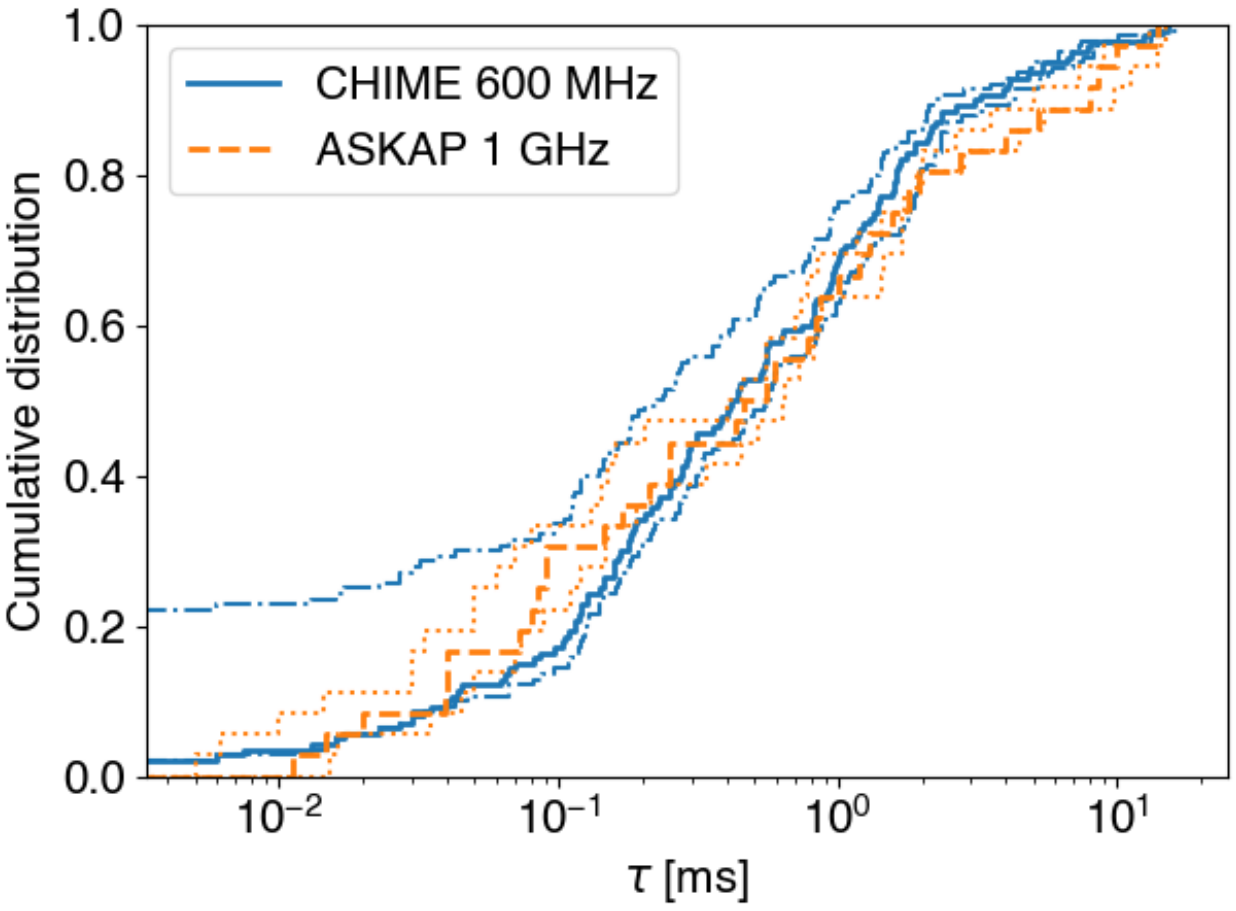

**Figure 4.** Comparison of CHIME (Sand et al., 2025) and ASKAP values of scattering. The upper and lower bounds are created by varying each FRB's scattering value by the quoted $1\sigma$ error. When an upper limit $\tau_{\rm max}$ only on scattering is quoted for CHIME, we use $\tau = 0.5\tau_{\rm max} \pm 0.5\tau_{\rm max}$.

Additional evidence comes from a lack of correlation between FRB scattering and redshift in our measurements, which was first hinted at through a lack of correlation between scattering and DM found by Sand et al. (2025). FRBs have their rest-frame scattering times suppressed by a factor of $(1+z)^3$ in the observer frame, so that any intrinsic peak in the scattering distribution is shifted to a lower apparent scattering time at higher redshifts. However, if the scattering distribution is approximately log-uniform, and dominated by an experimental cut-off, then the distribution will be largely redshift independent — as is observed.

Thus we suggest that no high-scattering downturn has been observed, and there is likely a much larger population of highly scattered FRBs.

#### 4.2.1 Multi-component structures mimicking scattering

Three FRBs (FRBs 20220725A, 20230526A, and 20230731A) show time-profiles that approximately resemble exponential scattering tails. However, all three have a 'break' in the exponential slope which is inconsistent with a single Gaussian plus exponential fit. Furthermore, all three initially begin as being 100% linearly polarised, but fall to 0% polarisation during the burst. Thus, they have been labelled as having 'ambiguous' structure in Table 1. We posit two scenarios: that these are true scattering tails, with the apparent break due to frequency dependence and/or statistical fluctuations, with the depolarisation due to either smearing of a variable PA and/or RM fluctuations in the screen; or that these FRBs consist of two distinct components — one polarised, one not — that mimic a scattered time profile.

FRB 20230526A has already been shown to exhibit scattering-induced depolarisation, consistent with varying RM through the scattering screen, though no evidence is found for either FRB 20220725A (Uttarkar et al., 2025) or FRB 20230731A (this work). While all three show variable PA during the polarised part of the burst, the PA varies smoothly and systematically over the entire polarised part of the burst until zero polarisation is reached, whereas a change in PA at the onset of depolarisation would be expected in the scattering scenario. FRB 20220725A also shows a circularly polarised component which is offset from the linear component, which would not be expected in the scattering explanation. Yet, our scattering fits find more scattering at lower frequencies in all three FRBs when fitting a single Gaussian plus scattering tail, with plausible values of $\alpha$ (-1.94 – -3.6). This does not in itself rule out multiple components, but does suggest that scattering is a significant contributor to the time-profile. For instance, fitting a second component to FRB 20230731A decreases the scattering timescale by only 20%, from 0.50 to 0.41 ms; we thus have quoted $\tau_{\rm obs} = 0.45 \pm 0.05$ ms. We thus posit that these FRBs truly have multiple components, but that the separation is comparable to the scattering timescale, resulting in an ambiguous identification.

Our sample also contains FRBs with partially overlapping components which are more clearly distinguished. FRB 20210320C has a second, elongated component which overlaps the primary peak, but with a clearly elongated tail that is clearly not consistent with an exponential fall-off due to scattering; while FRB 20230902A has a lower-amplitude secondary component which is fully resolved, but which might also mimic a scattering tail if the time-offset was smaller. In both cases however, the secondary bursts have identical polarisation properties to the primary. Nonetheless, we favour the multiple-component explanation.

Ultimately however, we can suggest no physical reason why depolarised components should only be present when mimicking a scattering tail, nor why depolarisation due to scattering should only operate on FRBs with poorly fit scattering tails. Until this phenomena is better understood, we urge caution in the interpretation of FRB scattering measurements, particularly for low S/N samples, since all three bursts would have been well-fit by a single scattering tail had their S/N not been so high.

#### 4.2.2 No correlation between scattering and excess dispersion measure

It has been suggested that the dominant cause of scattering in FRBs is the host galaxy's ISM, in which case the host ISM should impart an excess DM, and scattering should correlate with the host DM contribution, $\mathrm{DM}_{\rm Host,ISM}$ (for an extensive investigation into potential correlations between properties of this FRB sample and their host galaxies, see Glowacki et al., 2025). Thus, scattering could be used to improve redshift estimates for FRBs without identified hosts, and explain some of the variance in the Macquart relation (Cordes et al., 2022). Given that 29 of our FRBs have been localised to a host galaxy, we can estimate $\mathrm{DM}_{\rm Host,ISM}$ via

$$
\begin{aligned}
\mathrm{DM}'_{\rm Host} &\equiv \mathrm{DM}_{\rm FRB} - \mathrm{DM}_{\rm MW,ISM} \\
&\quad - \mathrm{DM}_{\rm MW,halo} - \mathrm{DM}_{\rm cosmic}, \qquad (9) \\
\mathrm{DM}_{\rm Host,ISM} &= (1+z)\mathrm{DM}'_{\rm Host} - \mathrm{DM}_{\rm Host,halo}. \qquad (10)
\end{aligned}
$$

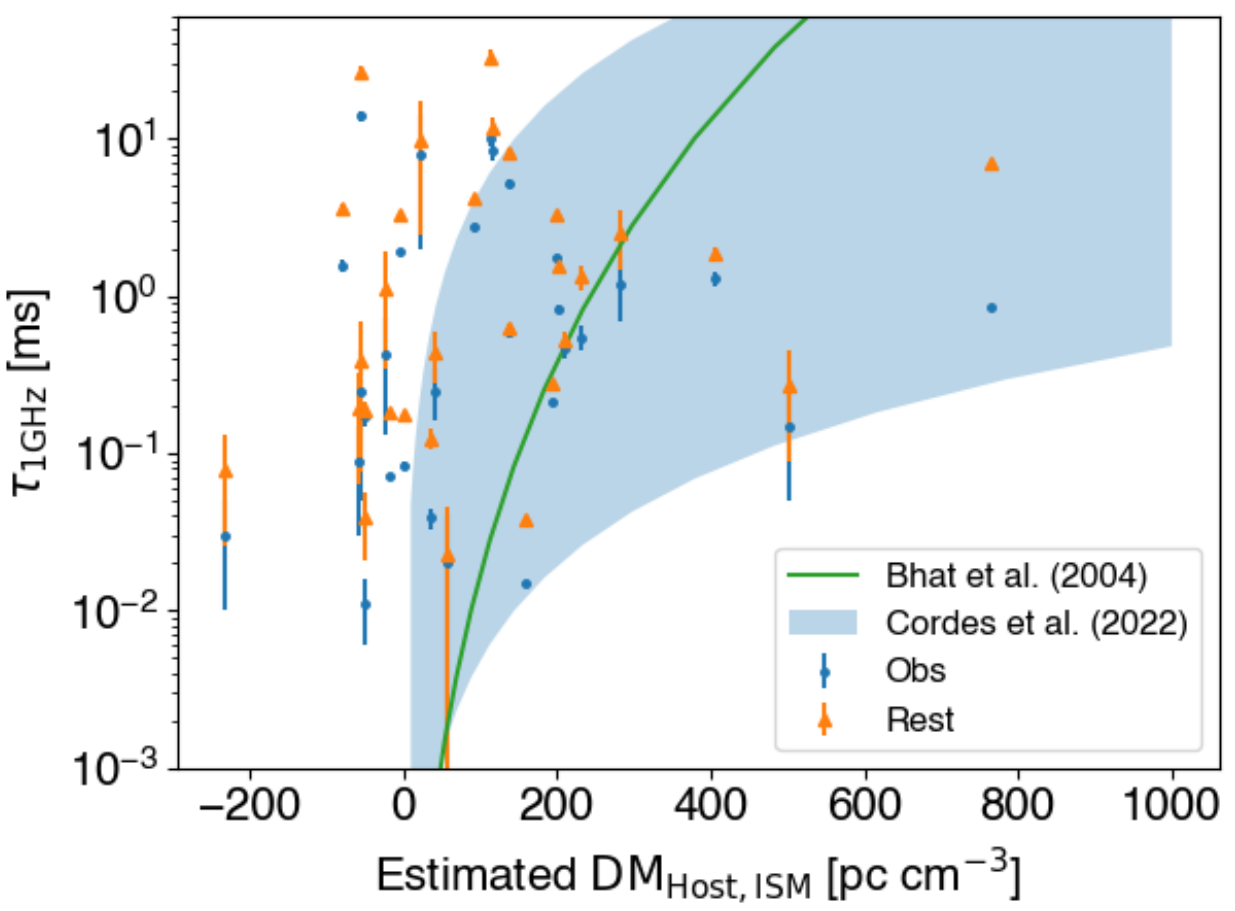

**Figure 5.** Plot of measured scattering times $\tau$, scaled to 1 GHz ('Obs') and to the host galaxy rest frame ('Host'), as a function of the estimated FRB host galaxy dispersion measures for localized FRBs in our sample. Also shown is the $\tau$–DM relation found for Galactic pulsars from Bhat et al. (2004), scaled by a factor of 3 to account for the Earth being at infinite distance from the FRB scattering screen; and the range of predictions for the 'cloudlet' model from Cordes et al. (2022), corresponding to scattering amplitude values of $0.01 \leq A_\tau \widetilde{F}G \leq 10$ (see that work for the meaning of these parameters).

Here, we assume contributions from the Milky Way's ISM according to the NE2001 model (Cordes & Lazio, 2003) with 50% error, as found by Schnitzeler (2012); we assume a halo contribution from the Milky Way of $\mathrm{DM}_{\rm MW,halo} = 50\,\mathrm{pc\,cm^{-3}}$ as per Prochaska & Zheng (2019); and we additionally assume a host halo $\mathrm{DM}_{\rm Host,halo} = 50\,\mathrm{pc\,cm^{-3}}$. The factor of $1 + z$ scales to the DM which would be measured in the rest-frame of the host galaxy. Note that Eq. 10 actually represents the deviation of the FRB DM from the Macquart relation, which is nominally attributed to total host contributions.

Figure 5 plots $\tau_{1\,\rm GHz}$ against the expression for $\mathrm{DM}_{\rm Host,ISM}$ in Eq. 10. There may be a hint of evidence for an upper limit on $DM_{\rm host,IGM}$ at low scattering amplitudes, consistent with the minimum amount of scattering expected from the cloudlet model of Cordes et al. (2022). However, over all the data, no correlation is evident, either by eye, or by fitting $\log_{10} \tau_{1\,\rm GHz}$ as a function of $z$, with or without adjustment to scattering in the host rest frame using a factor of $(1+z)^3$. As all FRBs in this sample bar one are observed at high Galactic latitudes ($|b| > 30^\circ$), we discount the influence of scattering in the Milky Way as obfuscating any underlying relation. The exception, FRB 20230718A, has values of $\tau_{1\,\rm GHz}$ (0.17 ms) and $\mathrm{DM}_{\rm Host,ISM}$ ($-48\,\mathrm{pc\,cm^{-3}}$) consistent with the bulk of the distribution.

There are several possible reasons for our observed lack of correlation. Firstly, the intrinsic variation in the DM–scattering relation observed for pulsars in the Milky Way is large: approximately a factor of three in DM for a fixed $\tau$ (equivalently, an order of magnitude in $\tau$ for a fixed DM; Bhat et al., 2004). However, even when adjusting for the expected three-fold increase in $\tau$ expected due to the Earth being at infinite distance from the scattering screen, many FRBs with low or negative estimated $\mathrm{DM}_{\mathrm{Host,ISM}}$ exhibit very large scattering times; while FRB 20240310A, with a very large estimated $\mathrm{DM}_{\mathrm{Host,ISM}} = 601.8\,\mathrm{pc\,cm^{-3}}$, shows very little scattering. Secondly, some of the variation in DM about the Macquart relation can be attributed to fluctuations in $\mathrm{DM}_{\mathrm{cosmic}}$ (Baptista et al., 2024), as recently found for FRB 20190520B by Lee et al. (2023). Thirdly, contributions from turbulent media in the vicinity of FRB progenitors may obey a different DM–$\tau$ relation to that of Galactic pulsars. Fourthly, the ISM sampled by FRBs traversing their host galaxy may be statistically different to the Milky Way ISM sampled by pulsars observed from Earth.

A Monte Carlo analysis by Mas-Ribas & James (2025) has recently shown that the first two explanations, combined with observational biases against observing highly scattered FRBs, would explain our lack of an observed correlation, even if the underlying $\tau$–DM relation in the host is well understood. Whichever the cause, we find that FRB scattering is not a good predictor of $\mathrm{DM}_{\mathrm{Host,ISM}}$ (i.e., the deviation of DM from the Macquart relation), and thus is unlikely to have utility as a predictor of the host galaxy redshift. We do not investigate, however, joint correlations with other estimators of DM, such as the $H_\alpha$ emission measure, as used by Ocker et al. (2022) in their analysis of FRB 20190520B.

#### 4.2.3 Scintillation

We find evidence for scintillation in 13 of our FRBs, with scintillation bandwidths in the range 0.11–9.2 MHz. Due to our narrow bandwidths, the fits to the scintillation index $\alpha_\nu$ are rarely consistent with the value $\sim$4 expected from scattering theory —- see Sammons et al. (2023) for an extended discussion. For a single scattering screen, we expect $2\pi\nu_{DC}\,\tau_{\mathrm{obs}} \equiv C \sim 1$ (Lambert & Rickett, 1999) — significant deviations from this indicates evidence for two (or more) scattering screens. We find evidence for two-screen scattering at greater than $2\sigma$ significance in six FRBs. Assuming that one screen exists in the Milky Way, and one is in the host galaxy, allows an upper limit to be placed on the distance product

$$L_x L_g \leq \frac{D_a^2}{2\pi\nu^2(1+z)}\frac{\nu_{\mathrm{DC}}}{\tau_{\mathrm{obs}}}, \tag{11}$$

where $L_g$ is the distance to the scattering screen in the Milky Way, $L_x$ is the distance from the FRB to the screen in the host, and $D_a \gg L_x, L_g$ is the angular diameter distance to the host galaxy. We calculate $D_a$ using host redshifts from Shannon et al. (2025) — available for five of these six FRBs — and standard Planck cosmology (Planck Collaboration et al., 2016), and calculate the maximum value, $(L_x L_g)_{\max}$, in Table 4.

Our results for the FRBs from Sammons et al. (2023) differ only in our estimates of $\tau_{\mathrm{obs}}$, which alters our quantitative estimates —- but not the qualitative conclusions — for FRBs 20190608B and 20210320C. Assuming a nominal Galactic screen distance $L_g \sim 1$ kpc, our new limits on the host scattering screens are not very constraining, allowing the host screens for FRBs 20230526A and FRB 20240210A to exist in their host halos, while the 'host' screen for FRB 20200906A could be at an extragalactic distance of 11.6 Mpc. We do not pursue more precise estimates of Galactic screen distances, which could be estimated from models such as NE2001 (Cordes & Lazio, 2003).

### 4.3 Polarisation properties

Our FRB sample exhibits a wide variety of polarisation properties, including temporal variation of the polarisation state within the bursts. A coherent transition from linear to circular polarisation is seen in FRB 20230708A where the total polarisation fraction remains high and roughly constant, while an incoherent transition between two orthogonal linear modes is apparent near the start of FRB 20221106A. FRB 20190611B's first component is fully linearly polarised, while its second component is evenly linearly/circularly polarised; thus, we cannot discern the nature of the transition. FRBs 20210912A and 20230731A exhibit circular polarisation time-profiles that are similar to, but offset from, their linear profiles. In particular, FRB 20210912A shows evidence within its primary pulse for an incoherent transition to a sub-pulse with differing linear PA; but the circular polarisation profile continues smoothly during this transition. FRBs 20220725A, 20230526A, and 20230731A show reducing polarisation fractions with time, the cause of which is discussed in detail in §4.2.1. Finally, FRBs 20190102C, 20220105A, and 20220918A also show evidence for changing polarisation behaviour, but the S/N of the relevant polarisation components is too low to discern the behaviour.

Here, we do not attempt a time-dependent analysis of polarisation fractions, and instead focus on the time-averaged FRB properties when discussing polarisation states.

#### 4.3.1 Lack of correlation between $L/I$ and $V/I$

Over our sample as a whole, we find the polarisation fractions L/I and V/I — integrated over both time and frequency — to be uncorrelated (see Figure 6), a result which was also found by DSA (Sherman et al., 2024). We therefore conclude that the observed circular polarisation is unlikely to have originated through conversion from linear polarisation in the same way for all FRBs. Note that in both §4.2.1 and §4.3.2, we disfavour a significant reduction in $L/I$ due to scattering-induced depolarisation. We emphasize that a proper investigation of the origin of circular polarization in individual FRBs requires a detailed study of the variation of polarisation state with time and frequency (e.g. McKinnon, 2024; Cho et al., 2020; Bera et al., 2025; Dial et al., 2025), which is beyond the scope of this paper and will be presented in a separate work.

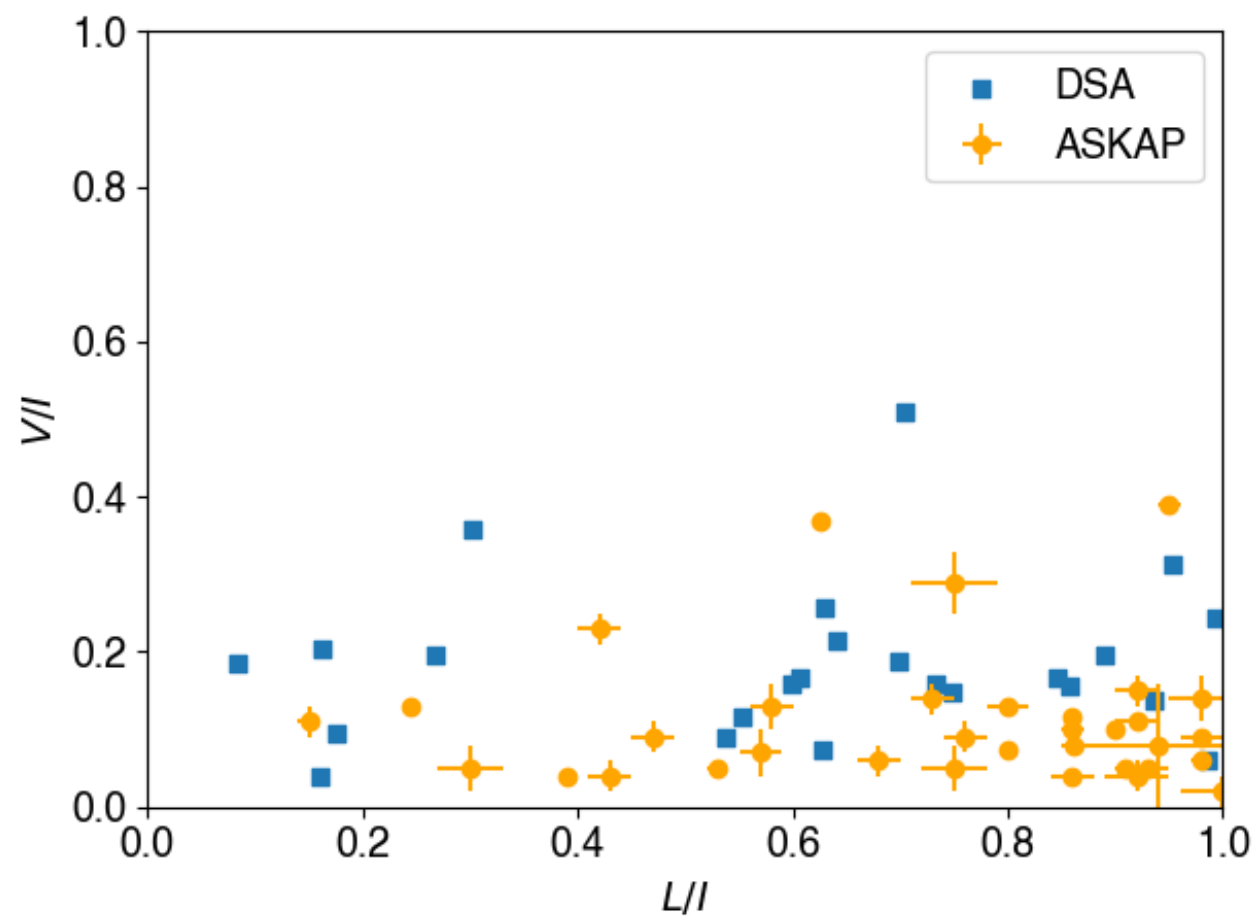

**Figure 6.** Circular vs. linear polarisation fractions for ASKAP FRBs, and those from DSA (Sherman et al., 2024).

### 4.3.2 *Position angle — macroscopic behaviour*

We divide the observed behaviour of PA into macroscopic and microscopic behaviour. The former is characterised by our polynomial fits to PA over the duration of the pulse, while the latter is characterised by short-term fluctuations about the overall trend.

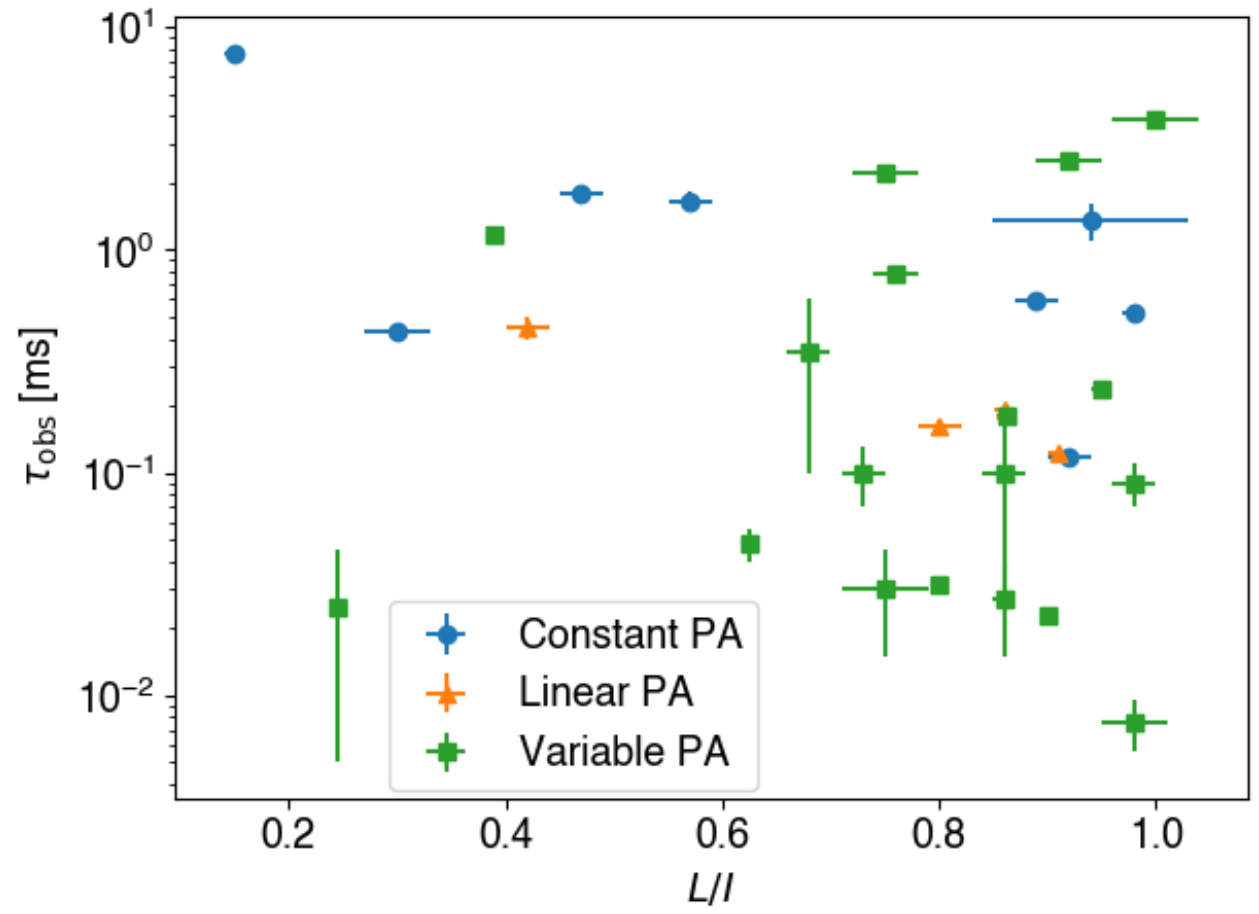

**Figure 7.** Scattering time at band centre, $\tau_{obs}$, as a function of linear polarisation fraction, L/I, for FRBs identified to have constant, linear, and variable PA behaviour.

Requiring a p-value of < 0.01 on an F-test to reject simpler FRB PA models, our polynomial fits find that 12 show constant, 4 show linear, and 17 exhibit variable trends, respectively. This confirms previous results that FRBs show a wide variety of PA behaviour (Sherman et al., 2024), which has previously been used as evidence for a magnetospheric origin of the emission (Luo et al., 2020). Pandhi et al. (2024) have performed a similar classification, using $\chi^2$ values to differentiate between constant and variable PA in 88 non-repeating FRBs detected by CHIME. Using a threshold $\chi^2$/n.d.f. > 5 to identify variable PA, those authors identify 19 FRBs with variable PA. If we use the same threshold, we identify only six FRBs with non-constant PA — consistent with the CHIME result. However, our higher average S/N allows us to resolve our FRBs on finer timescales, giving more independent measurements (i.e., degrees of freedom) with which to reject the hypothesis of a constant PA for a given $\chi^2$/n.d.f.. We expected that a higher degree of scattering seen at CHIME's lower frequencies may have also played a role, but as shown in §4.2, the observed distributions of scattering in both samples are similar.

There are many potential causes of PA variation. We do not see any evidence for FRBs with constant polarisation properties exhibiting a simple 'S'-shaped PA swing as expected from the rotating vector model (RVM) for pulsar emission (Hibschman & Arons, 2001), which has been observed in CHIME FRB 20221022A (Mckinven et al., 2025). However, two of our FRBs may exhibit such behaviour, but their polarisation behaviour is more complicated, as has been discussed elsewhere (Bera et al., 2024). Macroscopic PA behaviour may also be the result of more complex phenomenology (Beniamini & Kumar, 2025). Fitting of PA to such models will require accounting for the effects of scattering, which we leave to a future work.

Depolarisation by RM scattering is often associated with a PA swing due to the interference of many paths through a magnetised scattering screen. This has been observed in repeating FRBs (Feng et al., 2022), and a detailed analysis of 12 of the FRBs presented here has revealed such behaviour in FRB 20230526A (Uttarkar et al., 2025). However, as shown by Sand et al. (2025), there is no overall correlation between scattering and polarisation for FRBs. This is consistent with our sample — as Figure 7 shows, FRBs with high scattering are more likely to exhibit constant PA, and all FRBs with $\tau_{obs}$ below 0.01 ms have detectably variable PA, which is the opposite trend — and what is commonly observed in most pulsars (Li & Han, 2003). Furthermore, there is no correlation between PA behaviour and the degree of linear polarisation. This suggests that rather than scattering inducing PA swings, the dominant form of PA swing is intrinsic to the source, which scattering can then mask.

### 4.3.3 *Position angle — microstructure*

We find strong evidence for PA 'microstructure' — systematic and correlated variations in PA about the overall trend at timescales much shorter than the total burst duration. This is most evident using by-eye examination of the PA trend in strongly linearly polarised FRBs, where clear systematic behaviour at typically $\lesssim 0.1$ ms timescales is evident. We give an example from FRB 20240318A in Figure 8, which zooms in on two such features, showing systematic PA deviations of $\pm 20^\circ$ on $\sim 10\mu$s timescales — despite a nominal best-fit scattering timescale of 0.163 ms.

We can completely exclude that such effects are instrumental in origin. Such would apply on data prior to coherent dedispersion, and would be 'washed out' by the dispersion timescale, which is typically 0.1–1 seconds for the FRBs presented here — at least $10^3$ times longer than that of the observed microstructure.

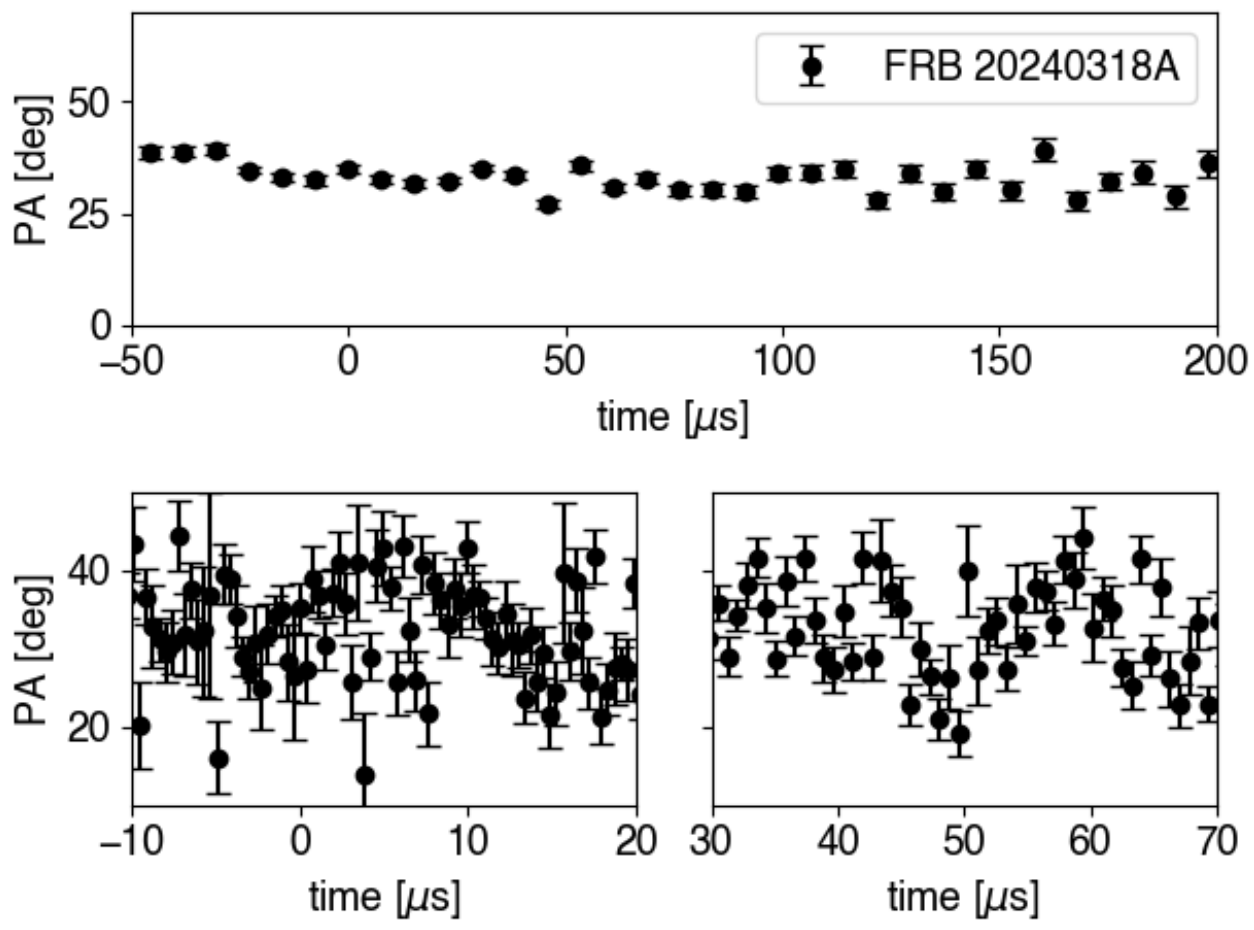

**Figure 8.** Example of PA microstructure in FRB 20240318A. Shown in the top panel is the PA integrated to 7.6μs, with zoom-ins at time resolutions of 0.38μs (bottom left) and 0.76μs (bottom right).

Rapid PA fluctuations ($\lesssim$ 100μ) by a few degrees in an FRB were first reported in a burst from FRB 20180916B by Nimmo et al. (2021) (see Figure 3, panel a, and Figure 4, panel c, of that work). Hewitt et al. (2023) analyse FRB 20220912A down to 1μs, and note "significant jumps" in the PA associated with microshots, all of which are approximately 100% linearly polarised. Such microstructure is however generally not visible in most FRB observations, since it requires a reliable polarisation calibration, high S/N, and high time resolution.

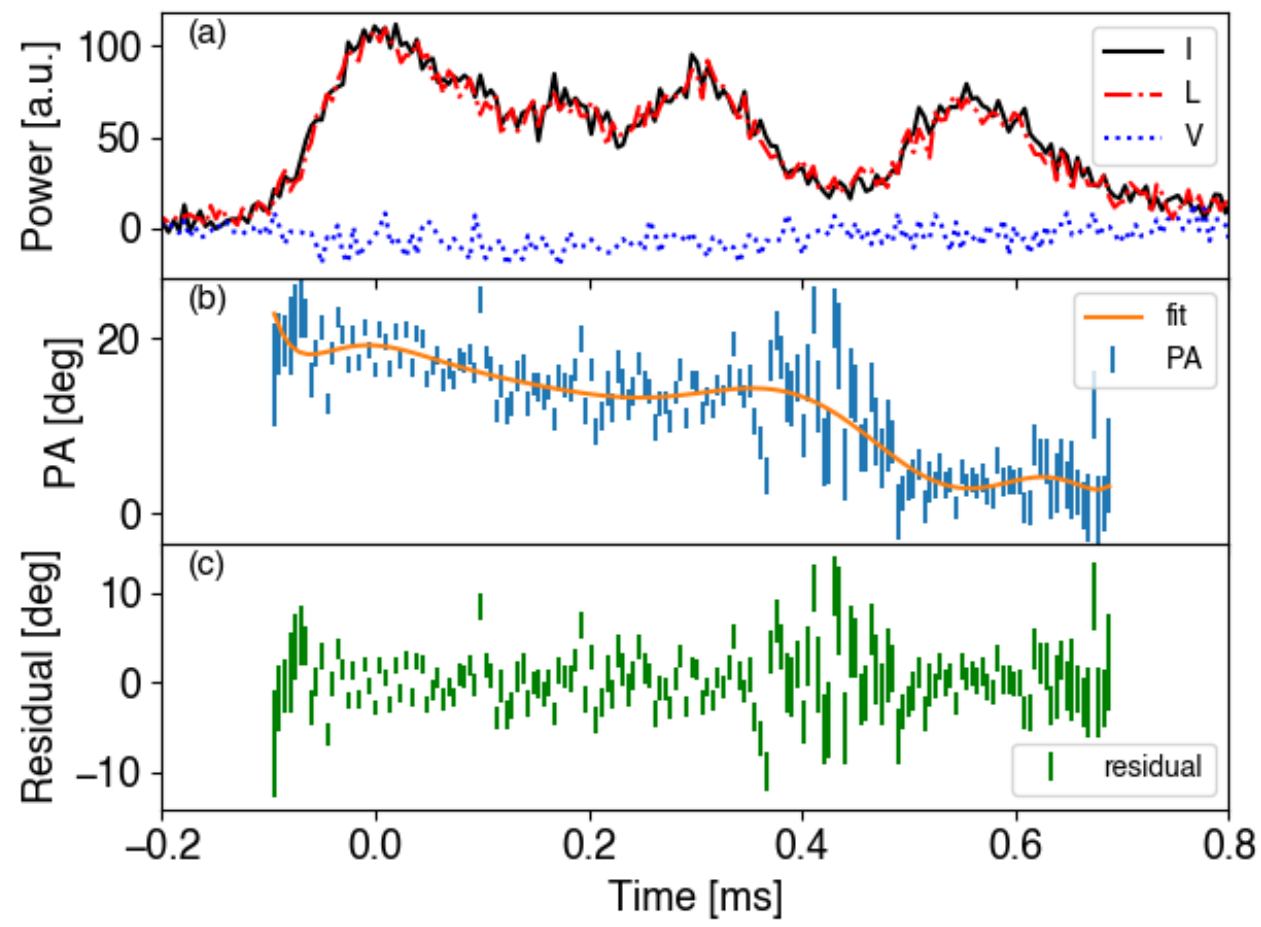

**Figure 9.** Example of a $10^{\text{th}}$ order polynomial fit to the PA in FRB 210407A, at an integration timescale of 4.95 μs. Panel (a): power of the polarisation components, (b): PA and polynomial fit, and (c): fit residuals.

Quantifying microstructure first requires integrating the Stokes parameters at timescales at which microstructure can be meaningfully observed, where too short an integration leads to noise-dominated data that resolves out the microstructure, while integrations that are too long smooth over it. For each FRB, we search all timescales between $10^{-3}w_{95}$ and $0.1w_{95}$, and use the criteria that total linear polarisation be four times its estimated error, $L > 4\sigma_L$ (noting that errors in $L$ are not Gaussian). We then place a cut on timescales yielding at least 20 such points. This yields 26 FRBs for which we can analyse microstructure.

The second step requires removing the macrostructure. Since the RVM does not appear to be generally applicable, we use a generic $10^{\text{th}}$ order polynomial fit to PA as a function of time. This allows fit residuals, $\epsilon_{\text{PA}}$, and an associated reduced chi-square value, $\chi^2$/n.d.f, to be calculated. We identify microstructure to be most significant at timescales producing the largest value of $\chi^2$/n.d.f, i.e. the worst fit to the trend. An example of this procedure is given in Figure 9.

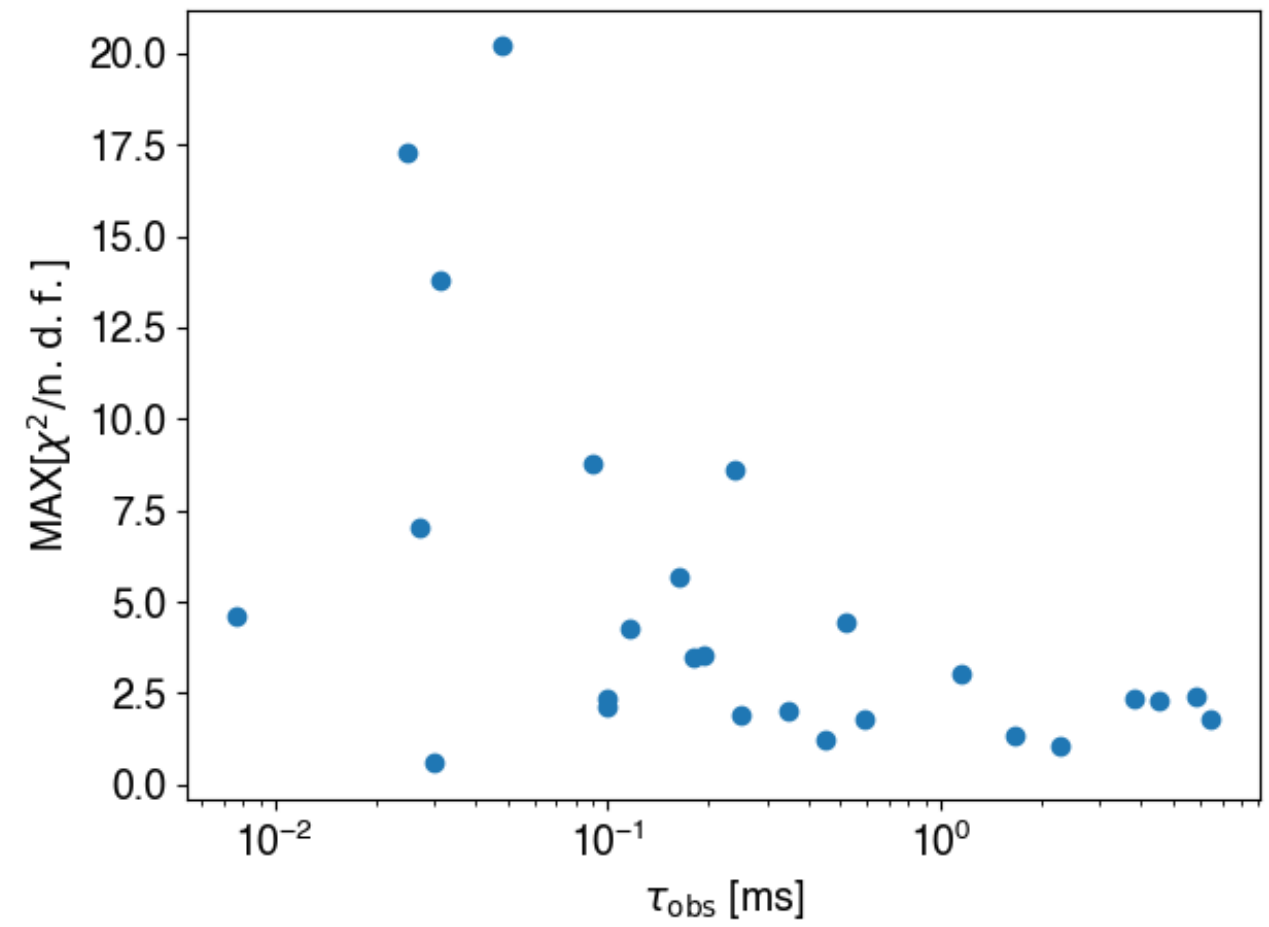

**Figure 10.** Correlation between PA microstructure and scattering: the maximum value of $\chi^2$/n.d.f against scattering timescale $\tau_{\text{obs}}$.

Performing this procedure, we find 13 FRBs with significantly high values of $\chi^2$/n.d.f, where we use a stricter probability cut of less than $10^{-4}$ (unadjusted for the trial factor) of the value occurring assuming that the polynomial fit is the true distribution. This occurs on timescales of 200μs to 200 ns, although this does not reliably identify a characteristic timescale for the microstructure. We find a clear correlation with our measured scattering timescale — while FRBs with low measured values of scattering may or may not show microstructure, no highly scattered FRBs do, as shown in Figure 10. Indeed, we suggest that effort be put into using these PA fluctuations to constrain scattering, given the ambiguities in scattering timescale from traditional modelling of Stokes I discussed in §4.2.1.

PA microstructure has long been known in pulsars (see e.g. Cordes, 1979), where single pulse phenomenology can be explained through the superposition of many sub-pulses, each with a different PA. This leads to a reduction in $L/I$ being correlated with PA swings, as the number of sub-pulses, and the variance in PA within sub-pulses, increases.

If PA microstructure is indeed associated with interference between sub-pulses, we expect reductions in $L$ to be associated with $\epsilon_{\text{PA}}$, and the number of interfering sub-bursts (van Straten, 2009). However, we have found no evidence for correlations between fluctuations in PA and decreases in $L/I$, with p-values of > 10% on Pearson's correlation coefficient being consistent with zero. Indeed, the 13 FRBs with significant microstructure have on-average a slightly higher linear polarisation fraction than those without, which we attribute to a selection effect — higher $L$ means greater resolution on PA. We can also exclude that our low-L/I FRBs are depolarised due to unresolved, random PA structure, since they exhibit systematic trends in their PA.

It is generally expected that such rapid PA variability requires the emission to arise within a neutron star magnetosphere, which is extremely hard to explain in models with emission arising far from the magnetosphere, and generally requires low radiative efficiency (Beniamini & Kumar, 2020).

The observed behaviour is thus most similar to that seen in a minority of Crab main pulses, which usually show random PA variation on nanosecond timescales, but sometimes produce systematic fluctuations on microsecond scales (Hankins et al., 2016). Unlike Crab main pulses however, our FRBs generally show little circular polarisation. Thus, the PA microstructure may be tracing turbulence down to $\sim$ 60 m resolution in the magnetic field of a parent body, likely a neutron star. It is remarkable to reflect that, given the Gpc distances to our FRBs, this is analogous to probing individual atoms on the outer planets.

We pause our exploration of FRB PA microstructure, and leave further investigation to a future work. We conclude only that it is not due to the interference of many unresolved bursts, and that likely it is ubiquitous, being unobserved primarily in FRBs with large scattering timescales.

## 4.4 Other properties

### 4.4.1 Multiple DMs

Our sample contains two FRBs — FRBs 20190611B and 20210407E — with distinct sub-components which appear, by-eye, to have a different dispersion measure to the rest of the burst. There are a further three FRBs — FRBs 20210117A, 20220501C, and 20230718A — where non-distinct sub-components may have a significantly different DM. In general, this could be due to intrinsic time-frequency burst properties, plasma or gravitational lensing, or a time-varying medium close to the FRB. Significant variations in DM, both between (Hessels et al., 2019) and within (Zhou et al., 2022) bursts, have previously been seen for repeating FRBs, and Sand et al. (2025) have also observed it in several once-off bursts, indicating that apparently once-off FRBs exhibit similar behaviour. We leave statistical tests searching for different DMs to a future work.

### 4.4.2 Repeater-like time-frequency profiles

The time-profiles of repeating FRBs tend to be broader in time and narrower in frequency (Pleunis et al., 2021), and may exhibit the 'sad-trombone' effect of frequency down-drifting of multiple components (Hessels et al., 2019; Josephy et al., 2019).

One of the FRBs in our sample — FRB 20190711A (Kumar et al., 2021) — has already been identified as a repeater, and clearly shows this structure. Additionally, FRBs 20220501C and 20231226A also exhibit frequency down-drifting, while FRB 20221106A shows no down-drifting, but has a repeater-like time-profile. This suggests these as potential candidates for follow-up observations.

### 4.4.3 Microshots in FRB 20190711A

We note remarkably short-scale structures in FRB 20190711A, the shortest and brightest of which is less than 50 μs in duration. This is paired with it being an overall broader FRB, with $w_{95}$ = 10.99 ms, among the largest in our sample. Its dynamic spectrum is qualitatively similar to several bursts from FRB 20220912A observed by Hewitt et al. (2023), being composed of short-scale "microshots" and broader-scale emission exhibiting a frequency downdrift over the total duration of the burst. Similar behaviour has been observed in FRB 20200120E by Majid et al. (2021), albeit with no frequency downdrift. Hewitt et al. (2023) observe residual (post-dedispersion) frequency drifts of the broader components of FRB 20220912A's bursts of a few hundred MHz ms$^{-1}$. While we cannot confidently assert with the S/N of our data that we observe similar frequency drifting within a single component of FRB 20190711A, the broad emission before the first microshot is qualitatively consistent with this possibility, though likely of order 30 MHz ms$^{-1}$.

### 4.4.4 Rotation measure

The RMs of our FRBs range from $-1711.8$ to $+1275$ rad m$^{-2}$. None exhibit the very large RMs of some FRBs (Michilli et al., 2018; Anna-Thomas et al., 2023), although several have RMs that are significantly larger than that found in our Milky Way (Hutschenreuter et al., 2022). We find no correlation between our FRB RMs and those expected from the MW, indicating either that FRB RMs are completely dominated by extragalactic contributions (most likely, turbulent plasma in the vicinity of their progenitor), or that the MW RM model is insufficiently fine-grained to apply to individual line of sight, or both. We expect an improvement in estimates for $\mathrm{RM}_{\mathrm{MW}}$ along our lines of site with the next data release from SPICE-RACS (Thomson et al., 2023), and from the Polarisation Sky Survey of the Universe's Magnetism (POSSUM; Gaensler et al., 2025).

## 5. Conclusions

We have presented new high-time resolution polarimetric measurements of 35 FRBs from the CRAFT ICS survey, made possible by ASKAP's real-time burst detection and voltage capture capabilities coupled with the CELEBI pipeline.

We have detected PA microstructure in the approximate range 200 μs to 200 ns in 13 of 26 FRBs with sufficient S/N. The high linear polarisation of bursts during these fluctuations excludes PA variation due to the superposition of multiple, independent nanosecond-scale sub-bursts as with Crab nanoshots, though it does resemble the PA behaviour of a minority of Crab main pulses.

Our analysis also shows that FRB scattering plays a major role in masking FRB substructure. Likely all FRBs are composed of multiple components, with sub-components being masked in highly scattered FRBs. Similarly, the probability of an FRB having identifiable variation in PA is strongly anti-correlated with the scattering timescale at band centre. We also find a generally smooth distribution of polarisation properties. Therefore, we suggest that previous classification schemes that use the number of identifiable components, variable or constant PA, and/or polarisation fractions result in arbitrary and artificial distinctions. We do, however, identify a sub-population of relatively highly circularly polarised FRBs ($|V| \geq 20\%$), which is also evident in DSA data, and encourage further studies of this sub-population.

Using the FRB localisations derived in Shannon et al. (2025), we have shown that FRB scattering is not detectably correlated with the deviation of an FRB's DM from the Macquart relation, and hence we do not recommend its use in host galaxy identification. We also cast doubt on our understanding of the FRB scattering distribution. We find that FRB sub-structure in Stokes I can mimic scattering, particularly at sub-ms timescales, and suggest that studying time-dependent polarisation properties may help to differentiate between true and apparent scattering tails, though we leave this to a future work. Similarly, a comparison of our scattering measurements with that of CHIME suggests that neither experiment has measured a maximum in the scattering probability distribution, and that a log-uniform distribution at high scattering values is a better fit than a log-normal.

We conclude by noting that the analysis presented here is generally not as detailed as the bespoke analyses presented in papers studying single FRBs, nor have we explored time-dependence in RM, DM, or frequency-dependence in polarisation properties. Rather, we have aimed to produce a uniform sample of FRBs with high-time-resolution polarised properties, the majority of which are also localised to their host galaxies. We encourage follow-up analysis using either derived properties presented in Tables 1–4, or using the Nyquist-resolution time–frequency data, available online.

**Acknowledgement**

This work was performed on the OzSTAR national facility at Swinburne University of Technology. The OzSTAR program receives funding in part from the Astronomy National Collaborative Research Infrastructure Strategy (NCRIS) allocation provided by the Australian Government, and from the Victorian Higher Education State Investment Fund (VHESIF) provided by the Victorian Government.

This work was supported by software support resources awarded under the Astronomy Data and Computing Services (ADACS) Merit Allocation Program. ADACS is funded from the Astronomy National Collaborative Research Infrastructure Strategy (NCRIS) allocation provided by the Australian Government and managed by Astronomy Australia Limited (AAL).

This scientific work uses data obtained from Inyarrimanha Ilgari Bundara, the CSIRO Murchison Radio-astronomy Observatory. We acknowledge the Wajarri Yamaji People as the Traditional Owners and native title holders of the Observatory site. CSIRO's ASKAP radio telescope is part of the Australia Telescope National Facility (https://ror.org/05qajvd42). Operation of ASKAP is funded by the Australian Government with support from the National Collaborative Research Infrastructure Strategy. ASKAP uses the resources of the Pawsey Supercomputing Research Centre. Establishment of ASKAP, Inyarrimanha Ilgari Bundara, the CSIRO Murchison Radio-astronomy Observatory and the Pawsey Supercomputing Research Centre are initiatives of the Australian Government, with support from the Government of Western Australia and the Science and Industry Endowment Fund.

This research has made use of NASA's Astrophysics Data System Bibliographic Services

We thank J.X. Prochaska for comments on the manuscript.

**Funding Statement** CWJ acknowledges support by the Australian Government through the Australian Research Council Discovery Projects funding scheme (project DP210102103). RMS acknowledges support through ARC Future Fellowship FT190100155 and Discovery Project DP220102305. MG is supported by the Australian Government through the Australian Research Council's Discovery Projects funding scheme (DP210102103), and through UK STFC Grant ST/Y001117/1. MG acknowledges support from the Inter-University Institute for Data Intensive Astronomy (IDIA). IDIA is a partnership of the University of Cape Town, the University of Pretoria and the University of the Western Cape. For the purpose of open access, the author has applied a Creative Commons Attribution (CC BY) licence to any Author Accepted Manuscript version arising from this submission. MWS acknowledges support from the Trottier Space Institute Fellowship program.

**Competing Interests** The authors declare no competing interests.

**Data Availability Statement** Data from this work is available online at https://doi.org/10.25917/1rg2-c612. Python packages NUMPY (Harris et al., 2020), SCIPY (Virtanen et al., 2020), and MATPLOTLIB (Hunter, 2007) were used in the preparation of this work.

## References

Anna-Thomas, R., Connor, L., Dai, S., et al. 2023, Science, 380, 599
Ashton, G., Hübner, M., Lasky, P. D., et al. 2019, ApJS, 241, 27
Bannister, K. W., Deller, A. T., Phillips, C., et al. 2019, Science, 365, 565
Baptista, J., Prochaska, J. X., Mannings, A. G., et al. 2024, ApJ, 965, 57
Beniamini, P., & Kumar, P. 2020, MNRAS, 498, 651
—. 2025, ApJ, 982, 45
Bera, A., James, C. W., Deller, A. T., et al. 2024, ApJ, 969, L29
Bera, A., James, C. W., McKinnon, M. M., et al. 2025, ApJ, 982, 119
Bhandari, S., Gordon, A. C., Scott, D. R., et al. 2023, ApJ, 948, 67
Bhat, N. D. R., Cordes, J. M., Camilo, F., Nice, D. J., & Lorimer, D. R. 2004, ApJ, 605, 759
Brentjens, M. A., & de Bruyn, A. G. 2005, A&A, 441, 1217
CHIME/FRB Collaboration, Amiri, M., Andersen, B. C., et al. 2021, ApJS, 257, 59
—. 2024, ApJ, 969, 145
Cho, H., Macquart, J.-P., Shannon, R. M., et al. 2020, ApJ, 891, L38

Cordes, J. M. 1979, Australian Journal of Physics, 32, 9
Cordes, J. M., & Lazio, T. J. W. 2003, arXiv e-prints, astro
Cordes, J. M., Ocker, S. K., & Chatterjee, S. 2022, ApJ, 931, 88
Day, C. K., Deller, A. T., Shannon, R. M., et al. 2020, MNRAS, 497, 3335
Dial, T., Deller, A. T., Uttarkar, P. A., et al. 2025, MNRAS, 536, 3220
Everett, J. E., & Weisberg, J. M. 2001, ApJ, 553, 341
Farah, W., Flynn, C., Bailes, M., et al. 2018, MNRAS, 478, 1209
Feng, Y., Li, D., Yang, Y.-P., et al. 2022, Science, 375, 1266
Gaensler, B. M., Heald, G. H., McClure-Griffiths, N. M., et al. 2025, PASA, 42, e091
Geyer, M., Karastergiou, A., Kondratiev, V. I., et al. 2017, MNRAS, 470, 2659
Glowacki, M., Bera, A., James, C. W., et al. 2025, arXiv e-prints, arXiv:2506.23403
Hankins, T. H., Eilek, J. A., & Jones, G. 2016, ApJ, 833, 47
Harris, C. R., Millman, K. J., van der Walt, S. J., et al. 2020, Nature, 585, 357
Hessels, J. W. T., Spitler, L. G., Seymour, A. D., et al. 2019, ApJ, 876, L23
Hewitt, D. M., Hessels, J. W. T., Ould-Boukattine, O. S., et al. 2023, MNRAS, 526, 2039
Hibschman, J. A., & Arons, J. 2001, ApJ, 546, 382
Hotan, A. W., Bunton, J. D., Chippendale, A. P., et al. 2021, PASA, 38, e009
Hunter, J. D. 2007, Computing in Science and Engineering, 9, 90
Hutschenreuter, S., Anderson, C. S., Betti, S., et al. 2022, A&A, 657, A43
Josephy, A., Chawla, P., Fonseca, E., et al. 2019, ApJ, 882, L18
Karastergiou, A., Johnston, S., Mitra, D., Van Leeuwen, A., & Edwards, R. 2003, MNRAS, 344, L69
Kellermann, K. I., & Pauliny-Toth, I. I. K. 1969, ApJ, 155, L71
Kolmogorov, A. 1933, G. Ist. Ital. Attuari., 4, 83
Krishnakumar, M. A., Joshi, B. C., & Manoharan, P. K. 2017, ApJ, 846, 104
Kumar, P., Shannon, R. M., Lower, M. E., Deller, A. T., & Prochaska, J. X. 2023, Phys. Rev. D, 108, 043009
Kumar, P., Shannon, R. M., Flynn, C., et al. 2021, MNRAS, 500, 2525
Lambert, H. C., & Rickett, B. J. 1999, ApJ, 517, 299
Lee, K.-G., Khrykin, I. S., Simha, S., et al. 2023, ApJ, 954, L7
Li, X. H., & Han, J. L. 2003, A&A, 410, 253
Lorimer, D. R., Bailes, M., McLaughlin, M. A., Narkevic, D. J., & Crawford, F. 2007, Science, 318, 777
Lu, W., & Kumar, P. 2019, MNRAS, 483, L93
Luo, R., Wang, B. J., Men, Y. P., et al. 2020, Nature, 586, 693
Majid, W. A., Pearlman, A. B., Prince, T. A., et al. 2021, ApJ, 919, L6
Mas-Ribas, L., & James, C. W. 2025, arXiv e-prints, arXiv:2508.13317
McKinnon, M. M. 2024, ApJ, 973, 35
Mckinven, R., Bhardwaj, M., Eftekhari, T., et al. 2025, Nature, 637, 43
Michilli, D., Seymour, A., Hessels, J. W. T., et al. 2018, Nature, 553, 182
Morrison, I. S., Bunton, J. D., van Straten, W., Deller, A., & Jameson, A. 2020, Journal of Astronomical Instrumentation, 9, 2050004
Ng, C., Pandhi, A., Naidu, A., et al. 2020, MNRAS, 496, 2836
Nimmo, K., Hessels, J. W. T., Keimpema, A., et al. 2021, Nature Astronomy, 5, 594
Nimmo, K., Hessels, J. W. T., Kirsten, F., et al. 2022, Nature Astronomy, 6, 393
Ocker, S. K., Cordes, J. M., Chatterjee, S., et al. 2022, ApJ, 931, 87
Oswald, L. S., Johnston, S., Karastergiou, A., et al. 2023, MNRAS, 520, 4961
Pandhi, A., Pleunis, Z., Mckinven, R., et al. 2024, ApJ, 968, 50
Planck Collaboration, Ade, P. A. R., Aghanim, N., et al. 2016, A&A, 594, A13
Platts, E., Weltman, A., Walters, A., et al. 2019, Phys. Rep., 821, 1
Pleunis, Z., Good, D. C., Kaspi, V. M., et al. 2021, ApJ, 923, 1
Prochaska, J. X., & Zheng, Y. 2019, MNRAS, 485, 648
Prochaska, J. X., Simha, S., Almannin, et al. 2023, doi:10.5281/zenodo.8125230
Purcell, C. R., Van Eck, C. L., West, J., Sun, X. H., & Gaensler, B. M. 2020, RM-Tools: Rotation measure (RM) synthesis and Stokes QU-fitting, Astrophysics Source Code Library, record ascl:2005.003
Sammons, M. W., Deller, A. T., Glowacki, M., et al. 2023, Two-screen scattering in CRAFT FRBs, arXiv:2305.11477
Sand, K. R., Curtin, A. P., Michilli, D., et al. 2025, ApJ, 979, 160
Schnitzeler, D. H. F. M. 2012, MNRAS, 427, 664
Scott, D. R., Cho, H., Day, C. K., et al. 2023, Astronomy and Computing, 44, 100724
Shannon, R. M., Bannister, K. W., Bera, A., et al. 2025, PASA, 42, e036
Sherman, M. B., Connor, L., Ravi, V., et al. 2024, ApJ, 964, 131
Smirnov, N. 1948, Annals of Mathematical Statistics, 19, 279
Speagle, J. S. 2020, MNRAS, 493, 3132
Sutinjo, A. T., Scott, D. R., James, C. W., et al. 2023, ApJ, 954, 37
Thomson, A. J. M., McConnell, D., Lenc, E., et al. 2023, PASA, 40, e040
Thornton, D., Stappers, B., Bailes, M., et al. 2013, Science, 341, 53
Uttarkar, P. A., Shannon, R. M., Gourdji, K., et al. 2025, arXiv e-prints, arXiv:2503.19749
van Straten, W. 2009, ApJ, 694, 1413
Virtanen, P., et al. 2020, Nature Methods, 17, 261
Zhou, D. J., Han, J. L., Zhang, B., et al. 2022, Research in Astronomy and Astrophysics, 22, 124001

## Appendix 1. Tabulated FRB data

We present the tabulated FRB properties in Tables 1, 2, 3, and 4. These will also be made available online.

**Table 1.** Measured properties of the bursts in our sample, being the mean observation frequency $\overline{\nu}$, the number of antennas used in the observation $n_{\rm ant}$, real-time detection S/N S/N$_{\rm det}$, offline analysis boxcar S/N S/N$_{\rm off}$, S/N-maximising DM estimate $\mathrm{DM}_{\mathrm{S/N}}$, structure-maximising DM estimate $\mathrm{DM}_{\mathrm{struct}}$, S/N-maximising width $w_{\rm snr}$, width containing 95% of the fluence $w_{95}$, and redshifts, $z$.

| FRB | $\overline{\nu}$ [MHz] | $n_{\rm ant}$ | S/N$_{\rm det}$ | S/N$_{\rm off}$ | $\mathrm{DM}_{\mathrm{S/N}}$ [pc cm$^{-3}$] | $\mathrm{DM}_{\mathrm{struct}}$ [pc cm$^{-3}$] | $w_{\rm snr}$ [ms] | $w_{95}$ [ms] | $z^b$ |
|---|---|---|---|---|---|---|---|---|---|
| 20180924B | 1320.0 | 24 | 21.1 | 77 | $361.75^{+0.01}_{-0.02}$ | $361.74^{+0.2}_{-0.08}$ | 0.91 | 2 | 0.3214 |
| 20181112A[a] | 1297.5 | 11 | 19.3 | 220 | $589.265 \pm 0.001$ | - | $\sim 0.1$ | $\sim 1.2$ | 0.4755 |
| 20190102C | 1297.5 | 23 | 14.0 | 167 | $364.55^{+0.02}_{-0.01}$ | $364.55^{+0.02}_{-0.02}$ | 0.076 | 1.25 | 0.29 |
| 20190608B | 1271.5 | 25 | 16.1 | 41 | $340 \pm 1$ | $338.7^{+1}_{-0.8}$ | 4.95 | 10.8 | 0.1178 |
| 20190611B | 1271.5 | 20 | 9.5 | 27 | $322.4 \pm 0.1$ | $322.7^{+0.4}_{-0.2}$ | 0.076 | 1.592 | 0.378 |
| 20190711A | 1271.5 | 28 | 23.8 | 46 | $592 \pm 2$ | $587.74^{+0.03}_{-0.02}$ | 8.6 | 10.99 | 0.522 |
| 20190714A | 1271.5 | 28 | 10.7 | 52 | $504.7 \pm 0.3$ | $504.13^{+0.5}_{-0.3}$ | 0.86 | 2.99 | 0.2365 |
| 20191001A | 919.5 | 30 | 37.1 | 108 | $507 \pm 0.3$ | $507^{+0.6}_{-0.7}$ | 5.3 | 13.468 | 0.23 |
| 20191228B | 1271.5 | 21 | 22.9 | 74 | $297 \pm 1$ | $296 \pm 2$ | 7.8 | 13.596 | 0.2432 |
| 20200430A | 863.5 | 26 | 13.9 | 67 | $379.9^{+0.6}_{-0.5}$ | $379.6^{+0.7}_{-0.4}$ | 11 | 22.68 | 0.161 |
| 20200906A | 864.5 | 7 | 16.1 | 347 | $577.8 \pm 0.2$ | $577.81^{+0.01}_{-0.01}$ | 0.057 | 0.128 | 0.3688 |
| 20210117A | 1271.5 | 21 | 17.7 | 112 | $729.2^{+0.2}_{-0.1}$ | $729.1^{+0.5}_{-0.2}$ | 1.24 | 3.584 | 0.214 |
| 20210320C | 863.5 | 24 | 15.3 | 161 | $384.6 \pm 0.02$ | $384.59^{+0.03}_{-0.01}$ | 0.381 | 0.884 | 0.28 |
| 20210407E | 1271.5 | 15 | 19.1 | 131 | $1784.8 \pm 0.2$ | $1784.9^{+0.3}_{-0.4}$ | 0.743 | 1.62 | N/A |
| 20210912A | 1271.5 | 22 | 31.2 | 479 | $1233.72^{+0.01}_{-0.02}$ | $1233.7^{+0.03}_{-0.02}$ | 0.095 | 1.612 | N/A |
| 20211127I | 1271.5 | 24 | 37.9 | 340 | $234.83 \pm 0.08$ | $234.86^{+14}_{-0.04}$ | 0.229 | 0.483 | 0.046946 |
| 20211203C | 920.5 | 24 | 14.2 | 47 | $636.2 \pm 0.4$ | $635.16^{+1.05}_{-0.82}$ | 12.4 | 25.449 | 0.3439 |
| 20211212A | 1631.5 | 24 | 10.5 | 45 | $200^{+2}_{-1}$ | $200^{+4}_{-3}$ | 2.1 | 5.628 | 0.0707 |
| 20220105A | 1631.5 | 22 | 9.8 | 42 | $583 \pm 2$ | $581.5^{+3.6}_{-3.1}$ | 0.95 | 2.25 | 0.2785 |
| 20220501C | 863.5 | 23 | 14.8 | 79 | $449.6^{+0.2}_{-0.4}$ | $449.1^{+0.3}_{-0.2}$ | 6.1 | 6.9 | 0.381 |
| 20220610A | 1271.5 | 22 | 23.9 | 62 | $1458.1^{+0.3}_{-0.5}$ | $1457.6^{+0.8}_{-0.9}$ | 1.07 | 2 | 1.015 |
| 20220725A | 919.5 | 25 | 10.9 | 49 | $290.1^{+0.3}_{-0.4}$ | $290^{+0.2}_{-0.3}$ | 3.43 | 8.016 | 0.1926 |
| 20220918A | 1271.5 | 24 | 26.3 | 21 | $643^{+6}_{-5}$ | $660^{+20}_{-30}$ | 11.43 | 13.851 | 0.491 |
| 20221106A | 1631.5 | 21 | 19.7 | 133 | $343.2^{+0.7}_{-0.9}$ | $343 \pm 0.3$ | 5.33 | 6.895 | 2.044 |
| 20230526A | 1271.5 | 22 | 22.1 | 88 | $316.2 \pm 0.2$ | $316.1^{+0.3}_{-0.1}$ | 2 | 2.7 | 0.157 |
| 20230708A | 919.5 | 23 | 30.5 | 270 | $411.54^{+0.05}_{-0.04}$ | $411.52^{+0.08}_{-0.05}$ | 1.14 | 23.578 | 0.105 |
| 20230718A | 1271.5 | 22 | 10.9 | 104 | $476.67 \pm 0.09$ | $476.64^{+0.2}_{-0.1}$ | 0.55 | 0.695 | 0.035 |
| 20230731A | 1271.5 | 25 | 16.6 | 61 | $701.1 \pm 0.3$ | $700.73^{+0.4}_{-0.5}$ | 0.65 | 2.655 | N/A |
| 20230902A | 832.5 | 22 | 11.8 | 113 | $440.1 \pm 0.1$ | $440.166^{+0.03}_{-0.01}$ | 0.229 | 0.678 | 0.3619 |
| 20231226A | 863.5 | 22 | 36.7 | 96 | $329.9 \pm 0.1$ | $328.73^{+2.16}_{-0.58}$ | 5.3 | 9.72 | N/A |
| 20240201A | 920.5 | 24 | 13.9 | 63 | $374.5 \pm 0.2$ | $373.514 \pm 0.35$ | 3.05 | 3.901 | 0.042729 |
| 20240208A | 863.5 | 14 | 12.1 | 21 | $260.2 \pm 0.3$ | $259.83 \pm 0.12$ | 1.7 | 10 | N/A |
| 20240210A | 863.5 | 23 | 11.6 | 59 | $283.73 \pm 0.05$ | $283.97 \pm 0.03$ | 0.305 | 1.539 | 0.023686 |
| 20240304A | 832.5 | 24 | 12.3 | 44 | $652.6 \pm 0.5$ | $653.4^{+5.2}_{-3.5}$ | 8.57 | 19 | N/A |
| 20240310A | 902.5 | 25 | 19.1 | 40 | $601.8 \pm 0.2$ | $601.76^{+0.86}_{-0.85}$ | 4.19 | 13.493 | 0.127 |
| 20240318A | 902.5 | 23 | 13.2 | 119 | $256.4 \pm 0.3$ | $256.18^{+0.02}_{-0.01}$ | 0.286 | 0.837 | N/A |

[a] Data from Cho et al. (2020).
[b] Data from Shannon et al. (2025).

**Table 2.** Measured scattering properties of the bursts in our sample. Given are the observed scattering times at band centre, $\tau_{\rm obs}$; the frequency dependence $\alpha$ ($\tau \sim \nu^{\alpha}$), the central frequency used for scattering fits $\overline{\nu}$, the fitted scattering value at 1 GHz, $\tau_{1\,\rm GHz}$, and whether or not the FRB apears, by-eye, to have a single ('s') component or multiple ('m') components; 'a' indicates ambiguity, discussed in §4.2.1.

| FRB | $\tau_{\rm obs}$ [ms] | $\alpha$ | $\overline{\nu}$ [MHz] | $\tau_{1\,\rm GHz}$ [ms] | Components |
|---|---|---|---|---|---|
| 20180924B | $0.59 \pm 0.01$ | $-3.68 \pm 0.04$ | 1297.5 | $1.56 \pm 0.13$ | s |
| 20181112A[a] | $0.023 \pm 0.002$ | $-2.0 \pm 0.3$ | 1297.5 | $0.039 \pm 0.006$ | m |
| 20190102C | $0.027 \pm 0.012$ | $-5.5 \pm 1$ | 1271.5 | $0.09 \pm 0.06$ | m |
| 20190608B | $3.83 \pm 0.15$ | $-3.37 \pm 1.3$ | 1269.5 | $8.5 \pm 1.2$ | s |
| 20190611B | $0.03 \pm 0.015$ | $0.3 \pm 2.0$ | 1252.7 | $0.03 \pm 0.02$ | m |
| 20190711A | $0.0076 \pm 0.002$ | $-2.5 \pm 1.1$ | 1172.9 | $0.011 \pm 0.005$ | m |
| 20190714A | $0.422 \pm 0.008$ | $-2.7 \pm 0.6$ | 1286.6 | $0.83 \pm 0.05$ | s |
| 20191001A | $4.52 \pm 0.03$ | $-4.85 \pm 0.3$ | 826.4 | $1.78 \pm 0.04$ | s |
| 20191228B | $5.85 \pm 0.2$ | $-3.6 \pm 0.6$ | 1273 | $14 \pm 1$ | s |
| 20200430A | $6.5 \pm 0.15$ | $-1.45 \pm 0.2$ | 863.5 | $5.25 \pm 0.25$ | s |
| 20200906A | $0.0315 \pm 0.0007$ | $-4.5 \pm 0.4$ | 846.4 | $0.0148 \pm 0.0004$ | s |
| 20210117A | $0.25 \pm 0.2$ | $5 \pm 8$ | 1274.5 | $0.15 \pm 0.1$ | m |
| 20210320C | $0.193 \pm 0.007$ | $-4.4 \pm 0.1$ | 828.4 | $0.084 \pm 0.004$ | m |
| 20210407E | $0.09 \pm 0.02$ | $-1.2 \pm 1.6$ | 1219.8 | $0.08 \pm 0.03$ | m |
| 20210912A | $0.048 \pm 0.008$ | $-2.5 \pm 0.9$ | 1275.6 | $0.09 \pm 0.03$ | m |
| 20211127I | $0.025 \pm 0.02$ | $0 \pm 5.5$ | 1272.5 | $0.02 \pm 0.02$ | m |
| 20211203C | $1.66 \pm 0.16$ | $-9.7 \pm 2.4$ | 891.4 | $0.55 \pm 0.1$ | s |
| 20211212A | $1.8 \pm 0.1$ | $-2.8 \pm 2.3$ | 1490.8 | $8 \pm 6$ | s |
| 20220105A | $0.43 \pm 0.01$ | $-2 \pm 0.8$ | 1649.8 | $1.2 \pm 0.5$ | m |
| 20220501C | $0.35 \pm 0.25$ | $4 \pm 8$ | 864.5 | $0.43 \pm 0.3$ | m |
| 20220610A | $0.521 \pm 0.001$ | $-3.56 \pm 0.03$ | 1149.4 | $0.855 \pm 0.008$ | s |
| 20220725A | $2.29 \pm 0.05$ | $-1.94 \pm 0.06$ | 1149.4 | $1.95 \pm 0.05$ | a |
| 20220918A | $7.66 \pm 0.1$ | $-2.10 \pm 0.03$ | 1133.5 | $10.0 \pm 1.1$ | s |
| 20221106A | $0.182 \pm 0.006$ | $-0.7 \pm 1.3$ | 1649.6 | $0.25 \pm 0.09$ | s |
| 20230526A | $1.16 \pm 0.01$ | $-3.6 \pm 0.3$ | 1272.2 | $2.75 \pm 0.1$ | a |
| 20230708A | $0.24 \pm 0.02$ | $-2.84 \pm 0.4$ | 920.5 | $0.21 \pm 0.005$ | m |
| 20230718A | $0.117 \pm 0.005$ | $-1.6 \pm 0.7$ | 1272.2 | $0.17 \pm 0.02$ | m |
| 20230731A | $0.45 \pm 0.05$ | $-2.3 \pm 0.6$ | 1271.8 | $0.78 \pm 0.04$ | a |
| 20230902A | $0.123 \pm 0.002$ | $-2.55 \pm 0.08$ | 812.4 | $0.072 \pm 0.002$ | m |
| 20231226A | $0.1 \pm 0.07$ | $-1 \pm 3$ | 762.8 | $0.25 \pm 0.2$ | m |
| 20240201A | $0.78 \pm 0.04$ | $-3.9 \pm 0.5$ | 915.5 | $0.46 \pm 0.06$ | m |
| 20240208A | $1.35 \pm 0.25$ | $-2.7 \pm 2.1$ | 864.1 | $1.0 \pm 0.45$ | s |
| 20240210A | $0.10 \pm 0.03$ | $-3.6 \pm 0.3$ | 863.5 | $0.59 \pm 0.04$ | m |
| 20240304A | $2.51 \pm 0.12$ | $3.5 \pm 1.3$ | 877 | $4 \pm 0.5$ | s |
| 20240310A | $2.23 \pm 0.07$ | $-3.23 \pm 0.5$ | 846.4 | $1.30 \pm 0.13$ | s |
| 20240318A | $0.163 \pm 0.01$ | $-3.32 \pm 0.005$ | 920.5 | $0.128 \pm 0.005$ | m |

[a] Data from Cho et al. (2020).

**Table 3.** Polarisation properties of the bursts in our sample, being linear polarisation fraction $L/I$, circular polarisation fraction $V/I$, and total polarisation fraction $P/I$; fitted rotation measure, RM, and expected Milky Way RM, RM$_{\mathrm{MW}}$, taken from Hutschenreuter et al. (2022); the polarisation calibrator used; and the fitted macroscopic PA trend. FRB 20190714A could not have its polarisation properties determined, due to missing data, while no RM could be fit for FRB 20240304A.

| FRB | $L/I$ | $\lvert V\rvert/I$ | $P/I$ | RM [$\mathrm{rad\,m^{-2}}$] | RM$_{\mathrm{MW}}$ [$\mathrm{rad\,m^{-2}}$] | Calibrator | PA trend |
|---|---|---|---|---|---|---|---|
| 20180924B | $0.89 \pm 0.02$ | $0.09 \pm 0.02$ | $0.90 \pm 0.02$ | $17.3 \pm 0.8$ | $16.5 \pm 5.0$ | VELA | constant |
| 20181112A[a] | $0.92$ | $0.19$ | $0.94$ | $10.5 \pm 0.4$ | $16.2 \pm 5.9$ | VELA | variable |
| 20190102C | $0.96 \pm 0.02$ | $0.02 \pm 0.02$ | $0.97 \pm 0.02$ | $-106.1 \pm 0.9$ | $26.6 \pm 7.7$ | VELA | variable |
| 20190608B | $1 \pm 0.04$ | $0.02 \pm 0.02$ | $1 \pm 0.04$ | $353 \pm 1$ | $-24.4 \pm 13.3$ | VELA | variable |
| 20190611B | $0.74 \pm 0.04$ | $0.29 \pm 0.04$ | $0.8 \pm 0.04$ | $17 \pm 3$ | $29.0 \pm 10.8$ | VELA | variable |
| 20190711A | $0.98 \pm 0.03$ | $0.14 \pm 0.02$ | $0.99 \pm 0.03$ | $4 \pm 1$ | $19.4 \pm 6.5$ | VELA | variable |
| 20190714A[b] | N/A | N/A | N/A | N/A | $-10.7 \pm 4.8$ | VELA | N/A |
| 20191001A | $0.53 \pm 0.01$ | $0.05 \pm 0.01$ | $0.54 \pm 0.01$ | $51.1 \pm 0.4$ | $23.5 \pm 4.3$ | VELA | const |
| 20191228B | $0.93 \pm 0.02$ | $0.1 \pm 0.02$ | $0.94 \pm 0.02$ | $11.9 \pm 0.9$ | $18.2 \pm 6.1$ | VELA | const |
| 20200430A | $0.43 \pm 0.02$ | $0.04 \pm 0.02$ | $0.43 \pm 0.02$ | $195.3 \pm 0.7$ | $14.5 \pm 7.0$ | 2045-1616 | const |
| 20200906A | $0.8 \pm 0.005$ | $0.073 \pm 0.004$ | $0.804 \pm 0.005$ | $75.47 \pm 0.08$ | $30.3 \pm 19.8$ | VELA | variable |
| 20210117A | $0.92 \pm 0.02$ | $0.05 \pm 0.01$ | $0.92 \pm 0.02$ | $-45.8 \pm 0.7$ | $3.3 \pm 9.2$ | VELA | const |
| 20210320C[c] | $0.86 \pm 0.008$ | $0.117 \pm 0.006$ | $0.868 \pm 0.008$ | $288.8 \pm 0.2$ | $-2.8 \pm 5.7$ | 1644 | linear |
| 20210407E | $0.97 \pm 0.01$ | $0.09 \pm 0.01$ | $0.98 \pm 0.01$ | $-9.1 \pm 0.6$ | $-59.6 \pm 28.6$ | VELA | variable |
| 20210912A | $0.625 \pm 0.005$ | $0.370 \pm 0.005$ | $0.726 \pm 0.004$ | $6.0 \pm 0.5$ | $8.4 \pm 3.8$ | VELA | variable |
| 20211127I | $0.244 \pm 0.003$ | $0.129 \pm 0.003$ | $0.276 \pm 0.003$ | $-67 \pm 1$ | $-2.9 \pm 6.2$ | None[d] | variable |
| 20211203C | $0.57 \pm 0.02$ | $0.07 \pm 0.03$ | $0.58 \pm 0.02$ | $34.3 \pm 1.2$ | $-29.2 \pm 9.1$ | 1644 | constant |
| 20211212A[c] | $0.47 \pm 0.02$ | $0.09 \pm 0.02$ | $0.48 \pm 0.02$ | $21 \pm 7$ | $6.0 \pm 5.7$ | 1644 | constant |
| 20220105A | $0.3 \pm 0.03$ | $0.05 \pm 0.03$ | $0.3 \pm 0.03$ | $-1312 \pm 8$ | $3.9 \pm 1.5$ | 1644 | constant |
| 20220501C | $0.68 \pm 0.02$ | $0.06 \pm 0.02$ | $0.69 \pm 0.02$ | $35.5 \pm 0.3$ | $9.6 \pm 4.5$ | VELA | variable |
| 20220610A | $0.98 \pm 0.01$ | $0.065 \pm 0.007$ | $0.99 \pm 0.01$ | $217 \pm 2$ | $11.9 \pm 4.9$ | VELA | constant |
| 20220725A[c] | $0.58 \pm 0.02$ | $0.13 \pm 0.03$ | $0.6 \pm 0.02$ | $-26.3 \pm 0.7$ | $-190.7 \pm 49.8$ | VELA | N/A[e] |
| 20220918A | $0.15 \pm 0.01$ | $0.13 \pm 0.02$ | $0.19 \pm 0.02$ | $559 \pm 23$ | $14.6 \pm 9.8$ | VELA | constant |
| 20221106A | $0.862 \pm 0.008$ | $0.078 \pm 0.006$ | $0.865 \pm 0.008$ | $444 \pm 1$ | $34.7 \pm 11.4$ | VELA | variable |
| 20230526A | $0.391 \pm 0.008$ | $0.04 \pm 0.008$ | $0.393 \pm 0.008$ | $613 \pm 2$ | $9.7 \pm 6.1$ | VELA | variable |
| 20230708A | $0.95 \pm 0.01$ | $0.39 \pm 0.01$ | $1.03 \pm 0.01$ | $-7.5 \pm 0.4$ | $43.6 \pm 10.5$ | VELA | variable |
| 20230718A | $0.92 \pm 0.02$ | $0.11 \pm 0.01$ | $0.92 \pm 0.02$ | $243.1 \pm 0.6$ | $186.4 \pm 50.4$ | 1644 | constant |
| 20230731A | $0.42 \pm 0.02$ | $0.23 \pm 0.02$ | $0.48 \pm 0.02$ | $268 \pm 5$ | $213.6 \pm 67.3$ | None[d] | linear |
| 20230902A | $0.91 \pm 0.01$ | $0.05 \pm 0.01$ | $0.91 \pm 0.01$ | $164.8 \pm 0.2$ | $10.1 \pm 6.3$ | VELA | linear |
| 20231226A | $0.86 \pm 0.02$ | $0.04 \pm 0.01$ | $0.86 \pm 0.02$ | $428.4 \pm 0.3$ | $13.0 \pm 6.8$ | VELA | variable |
| 20240201A[c] | $0.76 \pm 0.02$ | $0.09 \pm 0.02$ | $0.76 \pm 0.02$ | $1275.0 \pm 0.4$ | $5.9 \pm 6.5$ | 1644 | variable |
| 20240208A | $0.94 \pm 0.09$ | $0.08 \pm 0.08$ | $0.94 \pm 0.09$ | $-73.7 \pm 1.4$ | $3.9 \pm 6.3$ | 1644 | constant |
| 20240210A[c] | $0.73 \pm 0.02$ | $0.14 \pm 0.02$ | $0.74 \pm 0.02$ | $-325 \pm 1$ | $0.8 \pm 3.4$ | VELA | variable |
| 20240304A | $0.92 \pm 0.03$ | $0.04 \pm 0.02$ | $0.92 \pm 0.03$ | $489.7 \pm 0.8$ | $2.4 \pm 5.1$ | 1644 | variable |
| 20240310A | $0.72 \pm 0.03$ | $0.09 \pm 0.03$ | $0.72 \pm 0.03$ | $-1709.2 \pm 1.1$ | $-4.8 \pm 7.4$ | VELA | variable |
| 20240318A | $0.8 \pm 0.02$ | $0.13 \pm 0.01$ | $0.81 \pm 0.02$ | $-48.5 \pm 0.3$ | $14.3 \pm 2.3$ | 1644 | linear |

[a] Data from Cho et al. (2020).
[b] Only a single polarisation was available for this FRB.
[c] These FRBs were detected far from beam centre in edge or corner beams, and may have an incorrect polarisation calibration.
[d] No calibration observations taken for these FRBs.
[e] Too few data were available to estimate the PA trend.

**Table 4.** Measured scintillation properties of the bursts in our sample, giving modulation index $m$, decorrelation bandwidth $\nu_{\rm DC}$, spectral index of modulation $\alpha_{\rm DC}$ ($\nu_{\rm DC} \sim \nu^{\alpha}$), the time–bandwidth scattering–scintillation product $2\pi\nu_{\rm DC}\tau_{\rm obs}$, and (where applicable) an upper limit on the product of screen distances $L_z$ and $L_g$ (see text). Only those FRBs for which $\nu_{\rm DC}$ could be fit are shown.

| FRB | $m$ | $\nu_c$ [MHz] | $\nu_{\rm DC}$ [MHz] | $\alpha_{\rm DC}$ | $2\pi\nu_{\rm DC}\tau_{\rm obs}$ | $(L_z L_g)_{\rm max}$ [kpc$^2$] |
|---|---|---|---|---|---|---|
| 20190102C[a] | 0.41 | 1271.5 | $0.6 \pm 0.3$ | 10 | $101 \pm 68$ | |
| 20190608B[a] | 0.78 | 1271.5 | $1.4 \pm 0.1$ | $5.8 \pm 0.5$ | $33700 \pm 2700$ | $6.6 \pm 0.5$ |
| 20190611B | 0.96 | 1271.5 | $1.5 \pm 0.2$ | $-2 \pm 1$ | $282 \pm 146$ | |
| 20190711A[a] | 0.64 | 1136.9 | $0.11 \pm 0.01$ | $-10 \pm 5$ | $5.2 \pm 1.5$ | |
| 20200906A | 0.92 | 1271.5 | $1.99 \pm 0.01$ | $3 \pm 1$ | $394 \pm 9$ | $11,600 \pm 300$ |
| 20210320C[a] | 0.83 | 824.2 | $0.91 \pm 0.03$ | $2 \pm 1$ | $480 \pm 28$ | $1150 \pm 70$ |
| 20211127I | 0.74 | 1271.5 | $2.88 \pm 0.09$ | $3.3 \pm 0.2$ | $450 \pm 360$ | |
| 20220501C | 0.44 | 863.5 | $9.2 \pm 0.5$ | $-1 \pm 2$ | $20,200 \pm 14,500$ | |
| 20221106A | 0.84 | 1631.5 | $2 \pm 1$ | $1.4 \pm 0.5$ | $2290 \pm 1150$ | |
| 20230526A | 0.81 | 1271.5 | $2.6 \pm 0.1$ | $-6 \pm 4$ | $18950 \pm 750$ | $63 \pm 2$ |
| 20240210A | 0.71 | 863.5 | $2.7 \pm 0.1$ | $1 \pm 2$ | $1700 \pm 500$ | $58 \pm 17$ |
| 20240318A | 0.8 | 919.5 | $4.1 \pm 0.2$ | $1 \pm 1$ | $4200 \pm 330$ | |

[a] Data from Sammons et al. (2023), with updated $\tau_{\rm obs}$ from this work.